\input amstex
\input xy
\input epsf
\xyoption{all}
\documentstyle{amsppt}
\document
\magnification=1200
\NoBlackBoxes
\nologo
\hoffset=0.4in
\voffset=0.7in
\def\r{\roman}

\def\Q{\bold{Q}}
\def\Z{\bold{Z}^+}
\def\R{\bold{R}}
\def\C{\bold{C}}

\def\N{\bold{N}}
\vsize16cm


\bigskip

\centerline{\bf ERROR--CORRECTING CODES}

\medskip

\centerline{\bf AND PHASE TRANSITIONS}

\bigskip

\centerline{{\bf Yuri I.~ Manin, Matilde Marcolli}}

\bigskip

{\bf Abstract.} The theory of error-correcting codes is concerned with 
constructing codes that optimize simultaneously transmission rate and
relative minimum distance. These conflicting requirements determine
an asymptotic bound, which is a continuous curve in the space of
parameters. The main goal of this paper is to relate the asymptotic 
bound to phase diagrams of quantum statistical mechanical systems.
We first identify the code parameters with Hausdorff and von Neumann
dimensions, by considering fractals consisting of infinite sequences of code
words. We then construct operator algebras associated to individual
codes. These are Toeplitz algebras with a time evolution for which the
KMS state at critical temperature gives the Hausdorff measure on the
corresponding fractal. We extend this construction to algebras associated
to limit points of codes, with non-uniform multi-fractal measures, and
to tensor products over varying parameters.

\bigskip

{\bf Contents.}

\medskip

0. Introduction: asymptotic bounds

1. Spoiling Lemma

2. Asymptotic bound: existence theorem and unsolved problems

3. Code fractals: rate and relative minimum distance as Hausdorff dimensions

4. Operator algebras of codes

5. Quantum statistical mechanics and Kolmogorov complexity

6. Functional analytic constructions for limit points

7. The asymptotic bound as a phase diagram.

\bigskip

\centerline{\bf 0. Introduction: asymptotic bounds.}

\medskip

{\bf 0.1. Notation.} The following notation is used throughout the
paper. {\it An alphabet} is a finite set $A$ of cardinality $q\ge 2$, {\it a code} is a subset $C\subset A^n,\, 
n=n(C)\ge 1.$ {\it Words of length $n$} are elements of $A^n$, they are generally denoted $(a_1,\dots ,a_n), a_i\in A$
and alike. Elements of $C$ are {\it code words.}

\smallskip
{\it The Hamming distance} between two  words $(a_i)$, $(b_i)$ is defined as 
$$
d((a_i),(b_i)):= \# \{i\in (1,\dots ,n)\,|\,a_i\ne b_i\}.
$$
{\it The minimal distance} $d=d(C)$ of the code $C$ is 
$$
d(C) := \r{min}\,\{d(a,b)\,|\,a,b\in C, a\ne b\}.
$$
Finally, we put
$$
k=k(C):=\r{log}_q\,\#C,\quad [k]=[k(C)]:=\r{integer\ part\ of}\ k(C),
$$
so that
$$
q^{[k]}\le \#C=q^k<q^{[k]+1}.
\eqno(0.1)
$$
The numbers $n,k,d$ and $q$ are called {\it parameters of $C$}, and a code
$C$ with such parameters is called an $[n,k,d]_q$--code. Notice that any
bijective map between two alphabets produces a bijection between the associated
sets of codes, preserving all code parameters.

\smallskip

Alphabet $A$ and code $C$ may be endowed with additional structures.
The most popular case is: $A=\bold{F}_q$, the finite field with $q$
elements, and $C$ is a linear subspace of $\bold{F}_q^n$.
Such codes are called {\it linear ones.} 

\smallskip

Codes are used to transmit signals as sequences of code words. 
Encoding such a signal may become computationally more feasible,
if the code is a structured set, such as a linear space.
During the transmission, code words may be spoiled by a random noise,
which randomly changes letters constituting such a word. The noise  produces some word in $A^n$
which might not belong to $C$.
At the receiver end, the (conjecturally) sent word must be reconstructed, for example, 
as   closest neighbor in $C$ (in Hamming's metric) of the received word.
This  decoding operation again might become more computationally feasible,
if  $A$ and $C$ are endowed with an additional structure.

\smallskip

If $k$ is small with respect to $n$,  there are relatively few code words,
and decoding becomes safer, but the price consists in the respective
lengthening of the encoding signal. The number $R=R(C):=k/n,\ 0<R\le 1,$ 
that measures the inverse of this lengthening, is called the
(relative) {\it transmission rate.}  If $d$ is small, there might be too many code words
close to the received word, and the decoding becomes less safe. The number
$\delta :=\delta (C)=d/n,\ 0<\delta \le 1,$ is called {\it the relative minimal distance} of $C$.

\smallskip

The theory of error--correcting codes is concerned with studying and constructing
codes $C$ that satisfy three mutually conflicting requirements:

\smallskip

{\it (i) Fast transmission rate $R(C)$.
\smallskip

(ii) Large relative minimal distance $\delta (C)$.

\smallskip

(iii) Computationally feasible algorithms of producing such codes, together with feasible algorithms of
encoding and decoding.}

\smallskip

As is usual in such cases, a sound theory must produce a picture of limitations,
imposed by this conflict. The central notion here is that of {\it the asymptotic bound},
whose definition was given and existence proved in [Man]. The next subsection is devoted
to this notion.

\medskip

{\bf 0.2. Code points and code domains.}  We first consider all 
$[n,k,d]_q$--codes $C$ with fixed $q>1$ and  varying $n,k,d.$ 
To each such code we associate the point 
$$
P_C := (R(C),\delta (C)) =(k(C)/n(C), d(C)/n(C))\in [0,1]^2.
$$
Notice that in the illustrative pictures below the $R$--axis is vertical,
whereas the $\delta$--axis is horizontal: this is the traditional choice.

\smallskip

Denote by $V_q$ the set of all points $P_C$, corresponding to 
$[n,k,d]_q$--codes with fixed $q$. Let $U_q$ be the set of limit points of $V_q$.

\smallskip

In the latter definition, there is a subtlety. Logically, it might happen 
that one and the same code point corresponds to an infinite family
of different codes, but is not a limit point. Then we would have a choice, whether to include such points to $U_q$ automatically or not. However, we will show below
(Theorem 2.10), that in fact two possible versions of definition
lead to one and the same $U_q$.

\medskip

{\bf 0.3. Asymptotic bound.} The main result about code domain is this:
{\it $U_q$ consists of all points in $[0,1]^2$ lying below the graph of a certain continuous
decreasing function denoted $\alpha_q$:}
$$
U_q=\{(R,\delta )\,|\,R\le \alpha_q(\delta \}.
\eqno(0.2)
$$
This curve is called {\it the asymptotic bound}. Surprisingly little is known about it:
only various lower bounds, obtained using statistical estimates and explicit
constructions of families of codes, and upper bounds, obtained by rather simple count.

\smallskip

In any case, this bound is the main theoretical result describing limitations imposed by
the conflict between transmission rate and relative minimal distance.

\medskip

{\bf 0.4. Asymptotic bounds for structured codes.} If we want to take into account
limitations imposed by the  feasibility of construction, encoding and decoding as well,
we must restrict the set of considered codes, say, to a subset 
consisting of linear codes, or else polynomial time constructible/decodable codes
etc. Linear codes produce the set of code points denoted
$V^{lin}_q$ and the set of its limit points denoted $U^{lin}_q$. The latter domain
admits a description similar to (0.2), this time with another
asymptotic bound $\alpha^{lin}_q$. Clearly, 
$$
\alpha_q^{lin}(\delta ) \le \alpha_q (\delta ),
$$
but whether this inequality is strict is seemingly unknown.

\smallskip

Adding the restriction of polynomial computability, we get in the same way
asymptotic bounds  $ \alpha_q^{pol}(\delta )$ and $\alpha_q^{lin,pol}(\delta )$,
which are continuous and decreasing and lie below  the previous two bounds:
see [ManVla] and [TsfaVla].  

\smallskip

Proofs of (0.2) and its analogs are based upon a series of operations that allow one
to  obtain from a given code a series of codes with {\it worse}
parameters: the so called {\it Spoiling Lemma(s).} They form the subject of the next
section.

\medskip

{\bf 0.5. Asymptotic bounds as phase transitions.} In view of (0.2),
a picture of the closure of $V_q$ would consist of the whole domain
under the graph  of $\alpha_q$ plus  a cloud of isolated code points
above it. In a sense, the best codes are (some) isolated ones: cf. our
discussion in 2.5 and 2.6 below.

\smallskip

This picture reminds us e.~g. of phase diagrams in physics,
say, on the plane {\it (temperature, pressure)}, and alike. One of the goals
of this paper is to elaborate on this analogy.

\smallskip

To this end, we give several interpretations of $R$ and $\delta$
as ``fractional dimensions'', fractal and von Neumann's ones.

\bigskip

\centerline{\bf 1. Spoiling Lemma}

\medskip

{\bf 1.1. Code parameters reconsidered.} For linear codes, $k$ is always an integer.  
For general codes, this fails.
One can define $U_q$  using any one of the numbers $k/n$, $[k]/n$. As is easily seen,
they provide {\it the same} asymptotic bound $R=\alpha_q(\delta):$
$(k_i/n_i,d_i/n_i)$ and $([k_i]/n_i,d_i/n_i)$ diverge or converge simultaneously and have the same limit.
Working with both $k$ and $[k]$, depending on the context, can be motivated as follows.
\smallskip
(i) $k$ supplies the precise  cardinality of $C$, and the precise
transmission rate, but allows code points with irrational coordinates.
This introduces unnecessary complications both in the study of computability
properties of the code domains and in the statements of spoiling lemmas.

\smallskip

(ii) $[k]$ gives only estimates for $\#C$, but better serves spoiling.
Moreover, in the eventual studies of computability properties
of the graph $R=\alpha_q(\delta )$, it will be important to approximate
it by points with rational coordinates, rather than logarithms.

\smallskip

Unless stated otherwise, we associate with an $[n,k,d]_q$--code $C$
the code point  $(R(C):= k/n, \delta (C):=d/n)$, and define the family $V_q$
and the set $U_q$ using these code points.

\medskip

{\bf1.1.1. Spoiling operations.} Having chosen a code $C\subset A^n$
and a pair $(f,i)$, $f\in Map\,(C,A)$, $i\in \{1,\dots ,n\}$, define three new codes:
$$
C_1=:C*_if\subset A^{n+1}:\  
$$
$$
(a_1,\dots ,a_{n+1})\in C_1\
\r{iff} \ (a_1,\dots ,a_{i-1},a_{i+1},\dots ,a_n)\in C\,,
$$ 
$$
\r{and}\
a_i=f(a_1,\dots ,a_{i-1},a_{i+1} \dots ,a_n)\,.
\eqno(1.1)
$$

\smallskip
$$
C_2=:C*_i \subset A^{n-1}:\  
$$
$$
(a_1,\dots ,a_{n-1})\in C_2\
\r{iff}\ \exists b\in A, \ (a_1,\dots ,a_{i-1},b,a_{i+1},\dots ,a_n)\in C.
\eqno(1.2)
$$

\smallskip
$$
C_3=:C(a,i) \subset C\subset A^{n}:\quad\quad  
(a_1,\dots ,a_{n})\in C_3\
\r{iff}\ a_i=a.
\eqno(1.3)
$$

In plain words: operation $*_if$ inserts the letter $f(x)$ between the $(i-1)$--th
and the $i$--th letters of each word $x\in C$; operation $*_i$ deletes
the $i$--th letter of each word, i.~e. projects the code to the remaining coordinates; 
and $(a,i)$ collects those words of $C$
that have $a$ at the $i$--th  place.

\medskip

Assume now that $C$ is linear. 
\smallskip
Then  $C*_if$ remains linear,
if $f:\,C\to A=\bold{F}_q$ is a linear function. Moreover,
$C*_i$ is always linear. Finally, $C(a,i)*_i$ is also linear for any $a$.

\smallskip

These remarks will be used in order
to imply that Corollary  1.2.1. below remain true if we restrict ourselves to linear codes.

\medskip

{\bf 1.2. Lemma.} {\it If $C$ is an $[n,k,d]_q$--code,
then the codes obtained from it by application of one of these operations
have the following parameters:

\smallskip

(i) $C_1=C*_if$:\quad $[n+1,k,d]_q$, if $f$ is a constant function.

\smallskip

($i^{\prime}$) $C_1=C*_if$:\quad $[n+1,k,d+1]_q$, if for each pair $x,y\in C$ with
$d(x,y)=d$, we have $f(x)\ne f(y).$

\smallskip

(ii) $C_2=C*_i$:\quad $[n-1,k,d]_q$, if for each pair $x,y\in C$ with
 $d(x,y)=d$, these points have one and the same
letter at the place $i$. 

Otherwise $[n-1,k,d-1]_q$.

\smallskip

(iii) $C_3:=C(a,i)$. In this case, for each $i$,
there exists such a letter $a_i\in A$ (perhaps, not unique)  that
$$ 
\#C(a_i,i)\ge q^{k-1}.
\eqno(1.4)
$$
Therefore, the code $C(a_i,i)*_i$ will have parameters in the following range:}
$$
[n-1,\  k-1\le k^{\prime}<k,\  d^{\prime}\ge d]_q.
\eqno(1.5)
$$
\smallskip

{\bf Proof.} The statements $(i), (i^{\prime})$ and $(ii)$ are straightforward. For $(iii)$, 
remark that for any fixed $i$, $C$ is the disjoint union of $C(a,i),\,a\in A.$ Hence
$$
\sum_{a\in A} \#C(a,i)= q^k
\eqno(1.6)
$$
and $\#A=q$ together imply (1.4) for at least of one of $C(a,i)$. 
Passing to $C(a_i,i)*_i$, we are deleting the $i$--th letter
of all code words, which is common for all of them, so that the
minimal distance does not change. But for subcodes of $C$ it may be only $d$
or larger.

\medskip

{\bf 1.2.1. Corollary} (Numerical spoiling). {\it  If there exists
a code $C$ with parameters $[n,k,d]_q$, then there exist also a
code with the following parameters:

\smallskip

(i) $[n+1,k,d]_q$  (always).

\smallskip

(ii) $[n-1,k,d-1]_q$ (if  $n>1,k>0$.)

\smallskip

(iii) $[n-1,  k-1\le k^{\prime} <k, d]_q$ (if  $n>1$, $k>1).$

\smallskip

The same remains true in the domain of linear codes.}

\medskip

{\bf Proof.} Lemma 1.2 (i) provides the first statement.

\smallskip
In order to be able to use Lemma 1.2 (ii) for the second statement, 
we must find  a pair of words at the distance $d$ in $C$,
that have  different letters at some place $i$.
This is always possible if $\#C\ge 2,\, n\ge 2.$

\smallskip

The case (iii) can be treated as follows. 

\smallskip

If $C$ can be represented in the form 
$C^{\prime}*_ia$ where $a$ denotes the constant function 
$x\mapsto a\in C$, then $C^{\prime}$ is an $[n-1,k,d]_q$--code.
More generally, take the maximal projection of $C$
(onto some coordinate quotient set $A^m$) 
that is injective on $C$ and therefore preserves $k,d$.
We will get an $[m,k,d]_q$--code with $n>m\ge 2$, because 
for $m=1$ we must have $0<k\le 1$, the case that we have excluded in (iii).
If we manage to worsen its parameters to $[m, k^{\prime},d]_q$, $k-1\le k^{\prime}<k$,
then afterwards using (i) several times, we will get  an $[n-1,k^{\prime},d]_q$--code.

\smallskip
Therefore, it remains to treat the case when $C$ cannot be represented in the form 
$C^{\prime}*_ia$. In this case,  in the sum (1.6) there are at least two
non--vanishing summands. Hence for the respective code
$C(a,i)$ satisfying (1.4), we have also
$$
q^{k-1}\le \#C(a_i,i)<q^k.
\eqno(1.7)
$$
Therefore
$$
[k(C(a,i)*_i)]=[k]-1.
\eqno(1.8)
$$
It might happen that $d(C(a_i,i))>d$. In this case we can apply to $C(a_i)*i$
several times (ii) and then several times (i).
\medskip

{\bf 1.3. Remark.}  In the next section, we will prove the existence of the asymptotic
bound using only the numerical spoiling results of Corollary 1.2.1.
Thus such a bound exists for any subclass of (structured) codes
stable with respect to an appropriate family of spoiling operations,
in particular, for linear codes. Computational feasibility of  spoiled codes
must in principle be checked separately, but it holds for usual formalizations
of polynomial time computability.
\bigskip

\centerline{\bf 2. Asymptotic bound: existence theorem and unsolved problems}

\medskip

{\bf 2.1. Controlling cones.} Let $P=(R_P,\delta_P)$ be a point
of the square $[0,1]^2$  with $R_P+\delta_P<1$. All points of 
$U_q$ belong to this domain $\Delta$.

\smallskip

For two points $P,Q$, denote by $[P,Q]$ the closed
segment of the line $l(P,Q)$ connecting $P$ and $Q$.

\smallskip

For $P\in \Delta$, consider two segments $[P, (1,0)]$ and $[P,(0,1)]$,
The part of $\Delta$, bounded by these two segments and the
diagonal $R_P+\delta_P=1$, will be called
{\it the upper (controlling) cone of P} and denoted $C(P)_{up}.$

\smallskip

Extend  $[P, (1,0)]$ (resp. $[P,(0,1)]$) from their common point $P$
until their first intersection points with $\delta$--axis (resp. $R$--axis).
Then $\Delta$ will be broken into four  parts: the upper cone $C(P)_{up}$, the {\it lower cone} 
$C(P)_{low}$ lying below the lines  $l(P, (1,0))$ and $l(P,(0,1))$,  
{\it the left cone} $C(P)_l$  and {\it the right cone} $C(P)_r$.
We agree to include into each cone 
two segments of its boundary issuing from $P$.

\smallskip

\quad\quad\quad\quad\quad\quad\quad\quad\quad\quad  \epsfysize=40mm \epsfbox{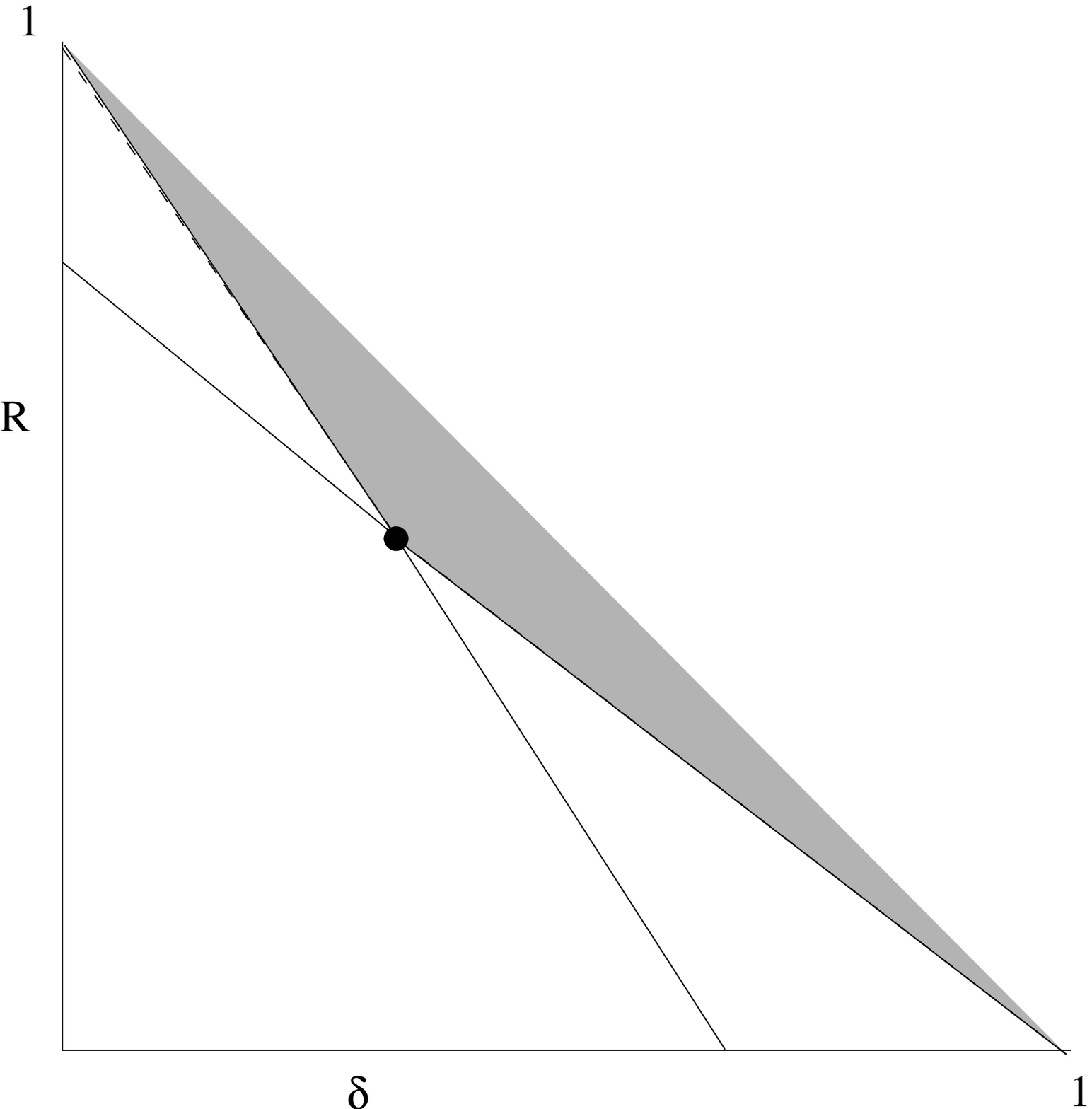}
\smallskip

\centerline{Fig.~1. \it Controlling cones}

\smallskip

Let $P,Q\in \Delta$.

\medskip

{\bf 2.1.1. Lemma.} {\it If $P\in U_q$, then  $C(P)_{low}\subset U_q$.}

\smallskip

This follows from the Spoiling Lemma. In the proof, it is convenient
to use the code points $([k]/n,d/n)$ rather than $(k/n,d/n)$.

\smallskip

In fact, if a sequence
of code points $Q_i=([k_i]/n_i,d_i/n_i)$ ($q$ being fixed) tends to
the limit point $(R,\delta )$, then the following statements are straightforward.

\smallskip

(a) $n_i\to \infty$.

\smallskip

(b) The boundaries of $C(Q_i)_{low}$ converge to the boundary
of $C((R,\delta))_{low}.$ Moreover,  the boundaries of $C(Q_i)_{low}$
contain code points that 
become more and more dense when $n_i\to \infty$ ,
namely $([k_i]-a)/n_i,d_i/n_i)$ and $([k_i]/n_i, (d_i-b)/n_i)$,
$a,b=1,2,\dots $ (Spoiling Lemma).

\smallskip

Thus,  the whole boundary of $C((R,\delta))_{low}$ belongs to
$U_q$.

\smallskip

(c) When a point $Q$ moves, say, along the right boundary segment
of $C((R,\delta ))_{low}$, the  left boundary segment
of $C(Q)_{low}$ sweeps the whole   $C((R,\delta ))_{low}$.

\smallskip

\quad\quad\quad\quad\quad\quad\quad\quad\quad  \epsfysize=40mm \epsfbox{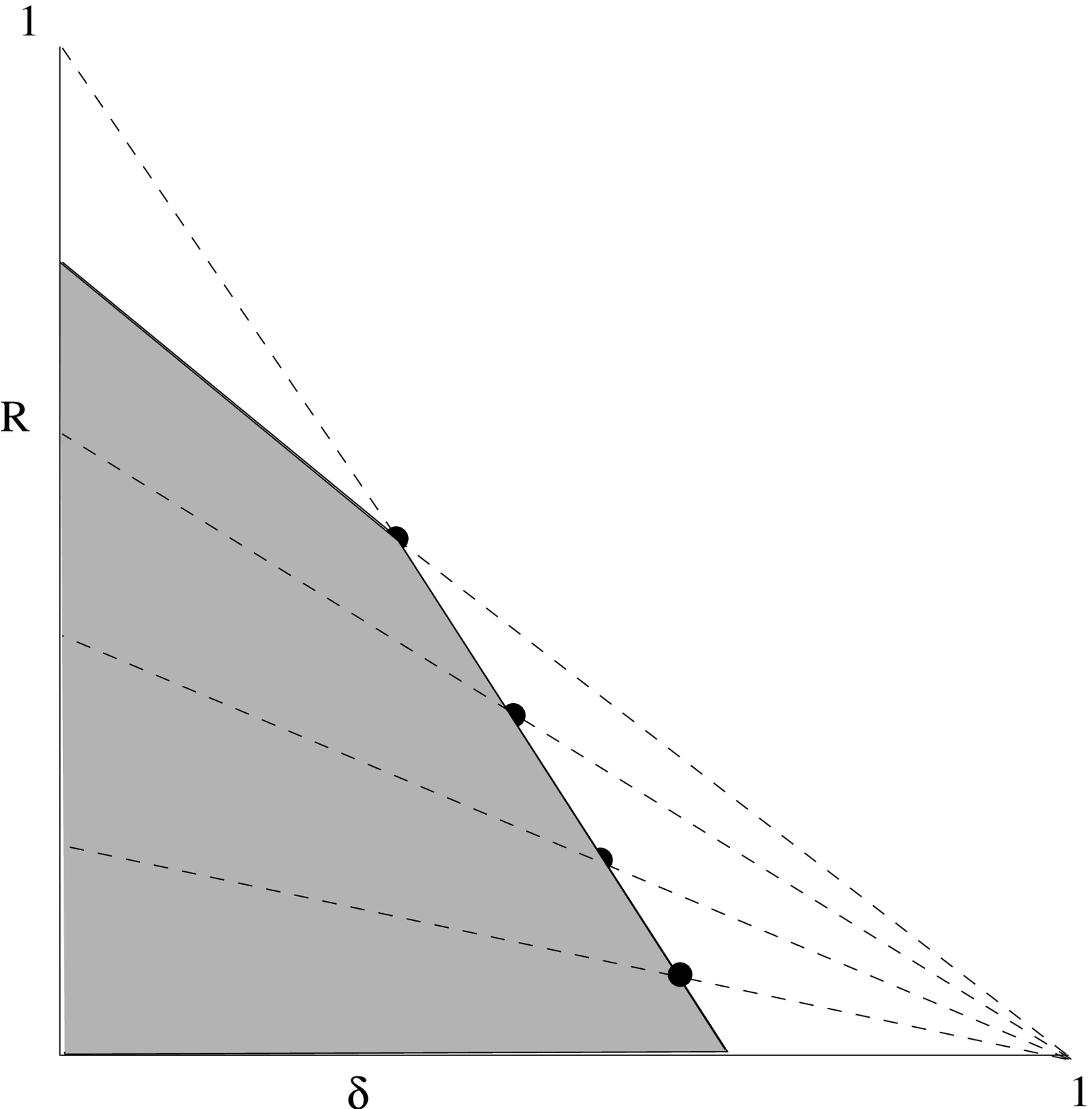}
\smallskip

\centerline{Fig. 2. {\it Code points on the lower cone boundary}}

\smallskip

This completes the proof of the Lemma. 

\medskip

{\bf 2.1.2. Lemma.} {\it (i) If $P\in C(Q)_l$, then $Q\in C(P)_r$,
and vice versa.
\smallskip

(ii) If $P\in C(Q)_{low}$, then $Q\in C(P)_{up}$,
and vice versa.}

\medskip

This is straightforward; a simple picture shows the reason.

\medskip 

{\bf 2.1.3. Lemma.} {\it If $P,Q\in\Gamma (\alpha_q)$  and $\delta_P<\delta_Q$,
then $P\in C(Q)_l$, and therefore $Q\in C(P)_r$.}

\smallskip
{\bf Proof.} In fact, otherwise  $P$ must be an inner
point of  $C(Q)_{low}$,  (or the same with $P,Q$ permuted).
But no boundary point of $U_q$ can lie in the lower cone
of another boundary point.

\medskip

{\bf 2.1.4. Controlling quadrangles.} Let $P,Q\in \Delta$, 
$\delta_P<\delta_Q$, and $P\in C(Q)_l$. Put
$$
C(P,Q):= C(P)_R\cap C(P)_l.
$$

\smallskip

\quad\quad\quad\quad\quad\quad\quad\quad\quad\quad  \epsfysize=40mm \epsfbox{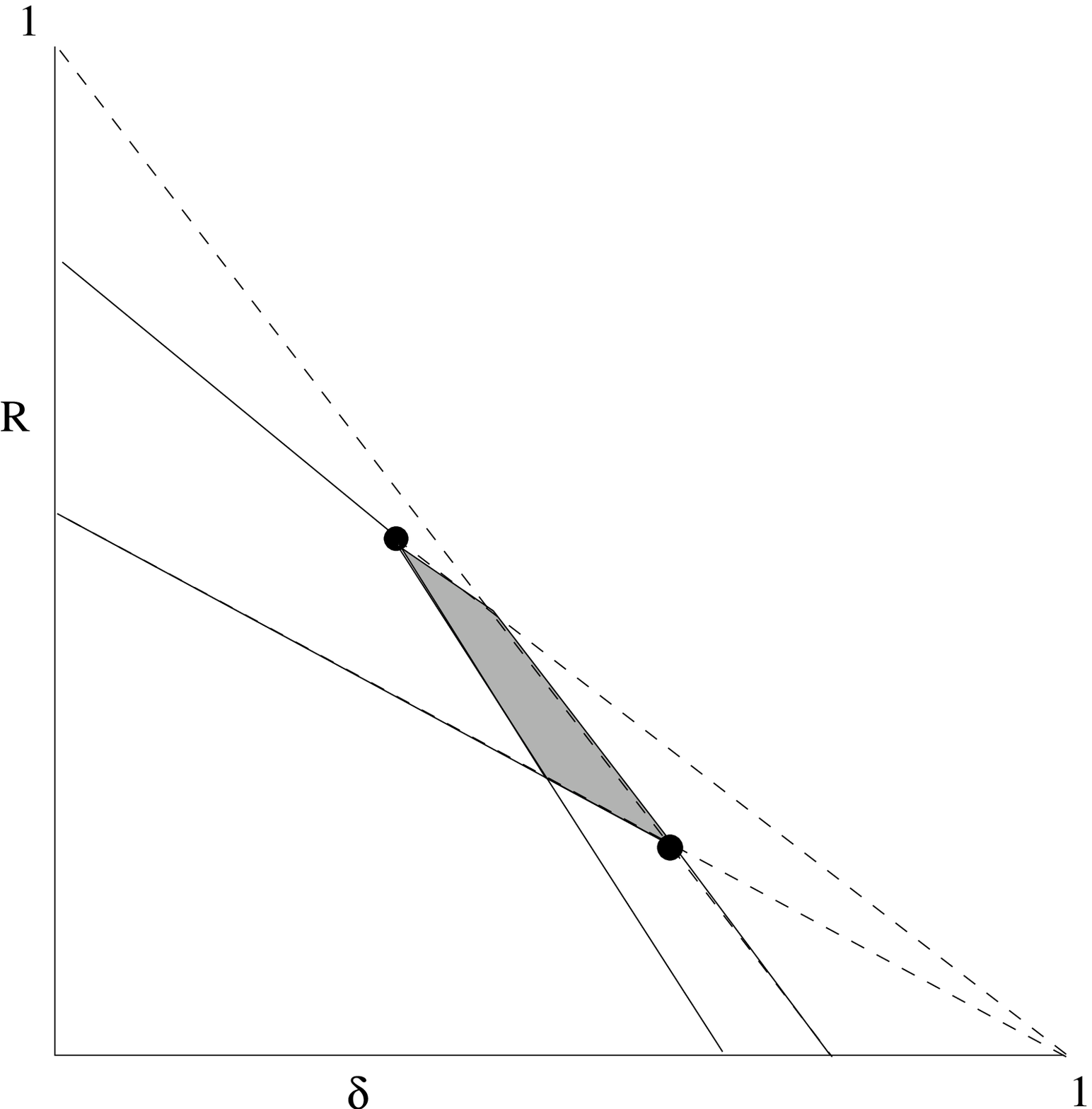}
\smallskip

\centerline{Fig. 3. {\it  Controlling quadrangle}}

\smallskip

When $P,Q\in \Gamma (\alpha_q)$ we will call $C(P,Q)$ the
{\it controlling quadrangle} with vertices $P,Q$.

\medskip

{\bf 2.1.5. Lemma.} {\it All points of $\Gamma (\alpha_q)$ between
$P$ and $Q$ belong to $C(P,Q)$.}

\smallskip

This follows from Lemma 2.1.3.

\smallskip

These facts suffice  to prove the following result ([Man], [ManVla]).

\medskip

{\bf 2.2. Theorem.} {\it For each $\delta \in [0,1]$, put
$$
\alpha_q(\delta ):= \r{sup}\, \{R_P\,|\,P = (R_P,\delta )\in U_q \}.
$$

Then 
\smallskip
(i) $\alpha_q$ is a continuous decreasing function. Denote its  graph by $\Gamma (\alpha_q)$.
We have $\alpha_q (0)=1$, $\alpha_q(\delta )=0$ for $\delta\in [(q-1)/q,1].$

\smallskip

(ii) $U_q$ consists of all points lying below or on  $\Gamma (\alpha_q)$. It is the union of all lower cones
of points of $\Gamma (\alpha_q)$.

\smallskip

(iii) Each horizontal line $0<R=const <1$ intersects $\Gamma (\alpha_q)$
at precisely one point, so that the $\Gamma (\alpha_q)$ is also the graph of the inverse function.

\smallskip

The same statement remains true, if we restrict ourselves by a subclass of
structured codes, for which Corollary 1.2.1 holds.}

\medskip

{\bf 2.3. Corollary.} {\it The curve $\Gamma (\alpha_q)$ (asymptotic bound)
is almost everywhere differentiable.}

\smallskip

This follows from the fact that it is continuous and monotone
(Lebesgue's theorem). 

\medskip

{\bf 2.4. Problem.} {\it (i) Is $\Gamma (\alpha_q)$ differentiable, or at least
peacewise differentiable?

\smallskip

(ii) Is this curve concave?} 

\medskip

{\bf 2.5. Isolated codes and excellent codes.}
Any code whose point lies strictly above
$\Gamma (\alpha_q)$ is called {\it isolated} one. Consider the union
$W_q$ of lower cones of all isolated codes.  This is a domain in $\Delta$
bounded from above by a piecewise linear curve, union of fragments of
bounds of these lower cones containing their vertices. A code is called 
{\it excellent} one, if it is isolated and is the vertex of one of
such fragments.

\medskip

{\bf 2.6. Problem.} {\it (i) Describe (as many as possible) excellent codes.

\medskip

(ii)  Are Reed--Solomon codes excellent in the class of linear, or even all  codes?}

\smallskip

Reed--Solomon codes  are certainly isolated, because they lie on the Singleton boundary
$R= 1-\delta +1/(q+1)$ which is higher than Plotkin's asymptotic bound 
$$
\alpha_q(\delta )\le 1- \delta -\frac{1}{q-1} \delta .
$$

\smallskip

One easily sees that the set of isolated points is infinite, and that 
points $R=1, \delta =0$ and the segment $R=0,
(q-1)/q \le \delta \le 1$ are limit points for this set.

\medskip

{\bf 2.7. Problem.} {\it Are there points on $\Gamma (\alpha_q)$,
$0<R<1$ that are limit points of a sequence of isolated codes?}

\medskip

{\bf  2.8. Code domain and computability.} The family $V_q$ is a 
recursive subfamily of $\Q$:  generating all codes and their
code points, we get an enumeration of $V_q$. Let $W_q:=\r{supp}\,V_q$
be the set of all code points.

\medskip

{\bf 2.8.1. Question.} {\it Is  $W_q$ a decidable set?}

\medskip

{\bf 2.8.2. Problem.} {\it Are the following  sets enumerable,
or even decidable:

\smallskip

(i) $\{(R(C),\delta(C))\,|\, R(C)< \alpha_q(\delta (C))\}$.

\smallskip

(ii) $\{(R(C),\delta(C))\,|\, R(C)\le \alpha_q(\delta (C))\}$.

\smallskip

(iii) $\{(R(C),\delta(C))\,|\, R(C)> \alpha_q(\delta (C))\}$.

\smallskip

(iv) $\{(R(C),\delta(C))\,|\, R(C)\ge \alpha_q(\delta (C))\}$.}

\medskip

{\bf 2.9. Codes of finite and infinite multiplicity.} Let $(R,\delta )$ be
the code point of a code $C$. We will say that this point
(and $C$ itself) has the finite (resp. infinite) multiplicity,
if the number of codes (up to isomorphism) corresponding to
this point is finite (resp. infinite). 

\smallskip
If $C$ has parameters $[n,k,d]_q$, then codes with the same code point
have parameters $[an,ak,ad]_q$, $a\in \Q^*_+$. Clearly,
finite (resp. infinite) multiplicity of $C$ can be inferred by looking at
whether there exist finitely or infinitely many $a\in \Q^*_+$
such that  an  $[an,ak,ad]_q$-code exists for such $a$.
Moreover, from the proof below one sees that one can restrict oneself by
looking only at integer $a$. 

\medskip

{\bf 2.10. Theorem.} {\it Assume that the code point of $C$
does not lie on the asymptotic bound. Then it has finite
multiplicity iff it is isolated.} 

\smallskip

{\bf Proof.}  If $C$ is of infinite multiplicity, it cannot be isolated. 
In fact, spoiling all codes with parameters $[an,ak,ad]_q$,
we get a dense set of points on the boundary of the lower cone
of the respective point.

\smallskip

Conversely, let an $[n,k,d]_q$--code $C$ lie below the asymptotic bound.
Then there exist $[N,K,D]_q$--codes with arbitrarily large $N,K,D$ satisfying the conditions
$$
\frac{K}{N} >\frac{k}{n}, \quad \frac{D}{N}>\frac{d}{n}.
\eqno(2.1)
$$
Slightly enlarging $N$ by spoiling, we may achieve $N=an$, with $a\in \N$.
Let
$$
K=ak^{\prime}+a_1,\  0\le a_1<a,\  k^{\prime}\in \N,
$$
$$
D=ad^{\prime}+a_2,\  0\le a_2<a,\  d^{\prime}\in \N,
$$
In view of (2.1), we have
$$
ak^{\prime}+a_1>ak,\quad  ad^{\prime}+a_2>ad.
$$
To complete the proof, it remains to reduce the parameters $K,D$
by spoiling, and get an 
$[an,ak,ad]_q$--code;
$a$ can be arbitrarily large.

\medskip

{\bf 2.11. Question.} {\it Can one find a recursive function
$b(n,k,d,q)$ such that if an $[n,k,d]_q$--code is isolated,
and $a>b(n,k,d,q)$, there is no code with parameters
$[an,ak,ad]_q$?}

\bigskip

\centerline{\bf 3. Code fractals: rate and relative minimum distance}
\centerline{\bf as Hausdorff dimensions}

\medskip

{\bf 3.1. Code rate and the Hausdorff dimension.} In this
subsection we will show that the rate $R$ of a code $C$ has a simple
geometric interpretation as the Hausdorff dimension of
a Sierpinski fractal naturally associated to the code.

\smallskip

We start with choosing a bijection of the initial alphabet $A$ with 
$q$--ary digits $\{0,1,\dots ,q-1\}$. Intermediary constructions will depend
on it, but basic statements will not. For the time being,
we will simply identify $A$ with digits.

\smallskip

The rational numbers with denominators $q^n$, $n\ge0$, admit
two different infinite $q$--ary expansions. Therefore we will exclude
them, and put 
$$
(0,1)_q:= [0,1]\setminus \{ m/q^n\,|\,m,n \in \bold{Z}\}
\eqno(3.1)
$$
The remaining points of the cube $x=(x_1,\dots ,x_n)\in (0,1)_q^n$ can be identified with
$(\infty\times n)$--matrices with entries in $A$: the $k$--th  column
of this matrix consists of the consecutive digits  of  the $q$--ary decomposition of $x_k$.

\smallskip

Now, for a code $C\subset A^n$, denote by $S_C\subset (0,1)_q^n$
the subset consisting of those points $x$, for which each line of the respective
matrix belongs to $C$. This is a Sierpinski fractal. 

\medskip

{\bf 3.2. Proposition.} {\it The Hausdorff dimension $s:=\r{dim}_H (S_C)$ equals to
the rate $R=R(C)$.}

\smallskip

{\bf Proof.}  $S_C$ is covered by $\#C=q^k$ cubes of size $q^{-1}$,
consisting of such points in $(0,1)^n$ that the first line of their coordinate matrix
belongs to $C$. Inside each such small cube lies a copy of $S_C$
scaled by $q^{-1}$. This self--similarity structure shows that $s$ is
the solution to the equation  $(\# C) q^{-ns} =1$ (see \S 9.2 of [Fal]).
Hence
$$
\dim_H(S_C) = \frac{\log(\# C)}{n \log q} = \frac{k}{n} = R.
\eqno(3.2)
$$
\medskip

{\bf Remark}. Several different notions of  fractal dimension
 (Hausdorff dimension, box counting
dimension, and scaling dimension) agree for $S_C$,  hence the
Hausdorff dimension can be computed from the simple
self--similarity equation. 
 
 \medskip

{\bf 3.3. Relative minimum distance and the Hausdorff dimension.} 
The most straightforward  way to connect the relative
minimum distance of a code $C$ with Hausdorff dimension is to
consider intersections  of $S_C$ with
$l$--dimensional linear subspaces $\pi =\pi^l$ that are
translates of intersections of coordinate hyperplanes in $\R^n$,
that is, are given by the equations $x_i=x_i^0$ for some
$i=i_1, \dots ,i_{n-l}$.    

\medskip

{\bf 3.3.1. Proposition.} {\it In this notation, we have: 

\smallskip

(i) If $l<d$, then $S_C\cap \pi$ is empty.

\smallskip

(ii) If $l\ge d$, then  $S_C\cap \pi$ has positive Hausdorff dimension:}
$$
\r{dim}_H(S_C \cap \pi)=\frac{\r{log}\, \#(C\cap \pi )}{l\,\r{log}\,q} >0.
\eqno(3.3)
$$

\smallskip

{\bf Proof.}  We will embed $C\subset A^n$ in $\R^n$ by sending
$(x_1,\dots ,x_n)$ to $(x_1/q,\dots ,x_n/q).$ (Notice that all
these points will lie in $[0,1]^n$, but outside of $(0,1)^n_q$.)

\smallskip

Then no two points of $C$ will lie in one and the same $l$--dimensional $\pi$,
if $n-l\ge n-d+1$, because at least $d$ of their coordinates are pairwise distinct. 
On the other hand, if $n-l\le n-d$, then one can find $\pi$ containing
at least two points of $C$. 

\smallskip

In terms 
of the iterative construction of the fractal $S_C$, this means the following. For a given 
$\pi$ with $l\leq d-1$, if the intersection $C\cap \pi$ is non--empty it must
consist of a single point. Thus, at the first step of the 
construction of $S_{C}\cap \pi$ we must replace the single cube $(0,1)^n_q\cap \pi$
with a single copy of a scaled cube of volume $q^{-l}$,
and then successively iterate the same
procedure. This will produces a family of nested open cubes of volumes 
$q^{-l N}$. Their  intersection is clearly empty.

\smallskip

When $l \ge d$, one can choose $\pi =\pi^d$ for which 
$C\cap \pi$ contains at least two points. Then the induced iterative construction of the
set $S_{C}\cap\pi$ starts by replacing the cube $Q^d =Q^n\cap \pi$ with $\# (C\cap \pi)$
copies of the same cube scaled down to have volume $q^{-d}$. The construction is then
iterated inside all the resulting $\#(C\cap \pi)$ cubes, so that one obtains a set 
of Hausdorff dimension $s=\dim_H(S_C\cap\pi)$ which is a solution
to the equation $\#(C\cap \pi)\cdot q^{-l s}=1$. Thus
$$
\r{dim}_H(S_C \cap \pi)=\frac{\r{log}\, \#(C\cap \pi )}{l\,\r{log}\,q} >0.
$$
This completes the proof.
\medskip

One can refine this construction by associating a fractal set $S_{\pi}$
to each subspace $\pi$ as above. Namely, define $S_{\pi}$
as the set of points of $(0,1)_q^n$ whose matrices have all rows
in $\pi$.

\medskip

{\bf 3.4. Proposition.} {\it The Hausdorff dimension of $S_\pi $ is
$$
\r{dim}_H S_{\pi} =\frac{l}{n}.
\eqno(3.4)
$$
In particular, for $l =d$ one has $\dim_H S_{\pi} =\delta$.}

\smallskip

{\bf Proof.} The argument is similar  to the one in the previous proof. 
We construct $S_\pi$ by subdividing, at
the first step, the cube $[0,1]^n$ into $q^n$ cubes of volume $q^{-n}$ and
of these we keep only those that correspond to points whose first
digit of the $n$-coordinates, in the $q$-ary expansion define a point
$(x_{11},\ldots,x_{1n})\in \pi \cap A^n$.  We have $\#(\pi \cap A^n)=q^{l}$,
hence at the first step we replace $Q^n$ by $q^l$ cubes of volume $q^{-n}$.
The procedure is then iterated on each of these. Thus, the Hausdorff
dimension of $S_\pi$ is the number $s$ satisfying $q^l q^{-ns}=1$, i.~e. (3.4). 

\medskip

One can now use $S_{\pi}$ in place of $\pi$, to make the roles
of rate and minimal relative distance more symmetric
in the Hausdorff context. Namely, we obtain,
\medskip

{\bf 3.5. Proposition.} {\it We have
$$
\r{dim}_H (S_C\cap S_{\pi})=\frac{\r{log}\,\# (C\cap \pi )}{n\,\r{log}\,q}
\eqno(3.5)
$$
In particular, for all $l \leq d-1$, the set $S_{C}\cap S_{\pi}$ is  empty. 
\smallskip
 For $l \geq d$,
there exists a subspace $\pi^l$ for which $\r{dim}_H (S_C\cap S_{\pi})>0$ so that
$S_C\cap S_{\pi}$ is a genuine fractal set.}

\smallskip

{\bf Proof.} Again, the argument is similar to the one we have already used.

\smallskip
The iterative construction
of $S_C \cap S_\pi$ replaces the initial unit cube $[0,1]^n$ with $\#(C\cap \pi)$ cubes
of volume $q^{-n}$ given by  points  with first row
$(x_{11},\ldots,x_{1n})\in C\cap \pi$. The same procedure is then iterated on each
of these smaller cubes. Thus, the Hausdorff dimension is given
by the self-similarity condition $\#(C\cap \pi)q^{-ns}=1$, which 
shows (3.5).
\smallskip
The same argument as above then shows that, for all
$l \leq d-1$ one has $\#(C\cap \pi)=1$, if $C\cap \pi$ is non-empty, while for
$l \geq d$ there exists a choice of $\pi$ for which $\#(C\cap \pi)\geq 2$.
This shows that once again $d$ is the threshold value for which
there exists a choice of $\pi \in \Pi_d$ for which $\dim_H(S_{C}\cap
S_{\pi})>0$.

\bigskip

\centerline{\bf 4. Operator algebras of codes.}

\medskip

{\bf 4.1. Finitely generated Toeplitz--Cuntz algebras.} 
We introduce a class of $C^*$--algebras related to codes.
Starting with an arbitrary finite set $D$, we associate to it
Toeplitz and Cuntz algebras, as in [Cu1], [Fow].

\medskip

{\bf 4.1.1. Definition} {\it (i) The Toeplitz--Cuntz algebra $TO_D$
is the universal unital $C^*$--algebra generated by a distinguished family of
isometries $T_d$, $d\in D$, with mutually orthogonal ranges.

\smallskip

(ii) The Cuntz algebra  $O_D$
is the universal unital $C^*$--algebra generated by a distinguished family of
isometries $S_d$, $d\in D$, with mutually orthogonal ranges, and
satisfying the condition}
$$
 \sum_{a\in D} S_d S_d^* = 1.  \eqno(4.1)
$$

\smallskip

Notice that $T_dT_d^*$ form pairwise orthogonal projections, so that operator
$$
P_D:= \sum_{a\in D} T_d T_d^* \in TO_D
$$
is a projector. But it is not identical.

\smallskip

From the definition it follows that the canonical morphism $TO_D\to O_D$:
$T_d\mapsto S_d$
generates the exact sequence
$$
0 \to J_D \to TO_D \to O_D \to 0,
$$
where $J_D$ is the ideal generated by $1-P_D$. The ideal $J_D$ is
isomorphic to the algebra of compact operators $\Cal{K}$.

\medskip

{\bf 4.1.2. Functoriality with respect to $D$.} The Toeplitz--Cuntz algebras
$TO_D$ are functorial with respect to arbitrary
{\it injective} maps $f:\,D\to D^{\prime}$: the respective
morphism maps $T_d$ to $T_{f(d)}$.

\smallskip

The Cuntz algebras are functorial only  with  respect to {\it bijections}: 
any bijection $f:\,D\to D^{\prime}$ generates an isomorphism $O_D\to O_{D^{\prime}}$
so that isomorphism class of $O_D$ depends only on $ \# D$.
The algebra $O_{\{1,\dots N\}}$ is often denoted simply $O_N$.

\smallskip

Below we will consider, in particular,  $TO_C$ and $O_C$ for codes $C$,
including codes $A^n$. The last remark allows us to canonically identify versions of $O_C$
that arise, for example, from different bijections $A\to \{0,\dots ,q-1\}$, as in 3.1
where they were used for the construction of fractals $S_C$.

\smallskip

Functoriality of $TO_D$ with respect to injections allows one to define
the algebra $TO_{\infty}:=TO_{\{1,2,\dots,\dots\}}$, see e.g. [Fow], identified 
with the algebra $O_{\infty}$ considered by Cuntz in [Cu1] and treated 
separately there.

\medskip

{\bf 4.1.3. Fractals and algebras.}  In order to connect Toeplitz--Cuntz and Cuntz algebras
$TO_C$, $O_C$ with fractals $S_C$, it is convenient to
 introduce two other topological spaces closely related to $S_C$.
 
 \smallskip
 
We will denote by $\bar S_C$ the closure
of the set $S_C$ inside the cube $[0,1]^n$, after identifying points of $S_C$
with $n$-tuples of irrational points in $[0,1]$ written in their 
$q$--ary expansion. The set $\bar S_C$ is also a fractal of the same Hausdorff 
dimension as $S_C$, which now includes also the rational points with $q$-ary
digits in $C$. It is a topological (metric) space in the induced topology from
$[0,1]^n$. 

\smallskip
We also consider the third space $\hat S_C$. It is  a compact
Hausdorff space, which maps surjectively to $\bar S_C$, one-to-one on $S_C$
and two-to-one on the points of $\bar S_C \smallsetminus S_C$.  By [Cu1] one
knows that $\hat S_C$ is the spectrum of the maximal abelian subalgebra of
the Cuntz algebra $O_C$. 

\smallskip

$\hat S_C$ can be identified with the set of all infinite
words $x=x_1 x_2 \cdots x_m \cdots$ with letters $x_i \in C$. 
Using the matrix language of 3.1, we can say that  points of $\hat C$
corresponds to all $(\infty ,n)$--matrices whose line belong to $C$.
The set $S_C$ is dense in $\hat S_C$ as the subset of non-periodic 
sequences.

\smallskip

The map $\hat S_C \to \bar S_C$ identifies coordinatewise the two $q$-ary expansions of
rational points with $q$--denominators in $\bar S_C$. The sets $\hat S_C$, $\bar S_C$ and $S_C$ only
differ on sets of Hausdorff measure zero, so for the purpose 
of measure theoretic considerations we often do not need to distinguish between them.

\smallskip

One can consider the abelian $C^*$-algebra $\Cal{A}(\hat S_C)$ generated by the characteristic
functions $\chi_{\hat S_C(w)}$, where $w=a_1 \cdots a_m$ runs over finite words with 
letters $a_i$ in $C$, and $\hat S_C(w)$ denotes the subset of infinite words $x\in \hat S_C$
that start with the finite word $w$. This algebra is isomorphic to the maximal abelian subalgebra
of $O_C$. In fact, these characteristic functions can be identified with the range projections
$P_w = S_w S_w^*=S_{a_1}\cdots S_{a_m} S_{a_m}^* \cdots S_{a_1}^*$ in $O_C$. 
We also denote by $T\Cal{A}(C)$ the abelian subalgebra of $TO_C$ generated by the
range projections $T_w T_w^*$, and which maps to $\Cal{A}(\hat S_C)$ in the quotient
algebra $O_C$.

\smallskip

Notice that, for an injective map $f: C \to C'$, the induced map $T_f: TO_C
\hookrightarrow TO_{C'}$ 
induces also an embedding $T_f: \Cal{A}(C) \hookrightarrow \Cal{A}(C')$
of the respective abelian subalgebras:
$$
T_w T_w^* \mapsto T_{f(w)} T_{f(w)}^* :=T_{f(a_1)}\cdots 
T_{f(a_m)} T_{f(a_m)}^* \cdots T_{f(a_1)}^* .
$$

\smallskip

For the sets $\hat S_C$ and the abelian algebras $\Cal{A}(\hat S_C)$, one
also has a functoriality in the opposite direction for more general maps
$f: C \to C'$ of codes that are not necessarily injective. Namely, such a
map induces a map $\hat S_C \to \hat S_{C'}$ that sends an infinite
sequence $x=a_1 a_2 \cdots a_m \cdots$ with $a_i\in C$ to the infinite 
sequence $f(a_1) f(a_2) \cdots f(a_m) \cdots$ in $\hat S_{C'}$. Since the
basis for the topology on $\hat S_C$ is given by the cylinder sets
$\hat S_C(w)$, the map constructed in this way is continuous. This
gives an algebra homomorphism $\Cal{A}(\hat S_{C'}) \to \Cal{A}(\hat S_C)$.

\medskip

{\bf 4.2. Representations of Cuntz algebras associated to $S_C$.}
In the following let us denote by $\sigma: S_C \to S_C$ the map that deletes the
first row of the coordinate matrix, shifting to the left the remaining $q$-adic 
digits of the coordinates,
$$
\sigma(x)=(x_{12}\dots x_{1k}\dots ; x_{22}\cdots x_{2k}\dots ;\dots ; x_{n2}\cdots x_{nk}\cdots) 
\eqno(4.2)
$$
for $x=(x_{11}x_{12}\cdots x_{1k}\cdots ; x_{21}x_{22}\cdots x_{2k}\cdots ;
x_{n1}x_{n2}\cdots x_{nk}\cdots)$ in $S_C$, that is, shifting upward the remaining
rows of the $\infty \times n$-matrix. For $a=(a_1,\ldots, a_n)
\in C\subset A^n$, let $\sigma_a$ denote the map adding $a$ as the first row of
the coordinate matrix
$$
\sigma_a(x)= (a_1 x_{11}x_{12}\dots x_{1k}\dots ; a_2 x_{21}x_{22}\dots x_{2k}\dots ; \dots ;
a_n x_{n1}x_{n2}\cdots x_{nk}\cdots).
\eqno(4.3)
$$
Since $a\in C$, (4.3) maps $S_C$ to itself. 
These maps are partial inverses of the shift (4.2). In fact, if we denote by
$R_a\subset S_C$ the range $R_a=\sigma_a(S_C)$, then on $R_a$ one has 
$\sigma_a \sigma(x)=x$, while for all $x\in S_C$ one has $\sigma \sigma_a(x)=x$.
We also introduce the notation
$$
\Phi_a(x) = \frac{d\mu\circ\sigma_a}{d\mu},
\eqno(4.4)
$$
for the Radon-Nikodym derivative of the Hausdorff measure $\mu$ composed with the map $\sigma_a$.
 
\medskip
 
Since the maps $\sigma_a$ act on $S_C$ by
$$
\sigma_a(x_1,\ldots,x_n) = \left( \frac{x_1+a_1}{q},\ldots, \frac{x_n+a_n}{q}\right),
\eqno(4.5)
$$
the Radon-Nikodym derivative $\Phi_a$ of (4.4), with $\mu$ the Hausdorff measure 
of dimension $s=\dim_H(S_C)$, is constant
$$
\Phi_a(x) = \frac{d\mu\circ\sigma_a}{d\mu} = q^{-ns} = q^{-k}.
\eqno(4.6)
$$

\medskip
 
{\bf 4.2.1. Proposition.} {\it
The operators
$$
(S_a f)(x) = \chi_{R_a} (x) \Phi_a(\sigma(x))^{-1/2} f(\sigma(x))
\eqno(4.7)
$$
determine a representation of the algebra $O_C$ on the Hilbert space $L^2(S_C,\mu)$.}

\smallskip

{\bf Proof.} The adjoint of (4.7) in the $L^2$ inner product $\langle \,,\rangle$  is of the form
$$
(S_a^* f)(x) = \Phi_a(x)^{1/2} f(\sigma_a(x)),
\eqno(4.8)
$$
therefore $S_a S_a^* = P_a$, where $P_a$ is the projection given
by multiplication by the characteristic function $\chi_{R_a}$, so that one obtains
$\sum_a S_a S_a^* =1$. Moreover, $S_a^* S_a =1$, so that one obtains a
representation of the $C^*$-algebra $O_C$.

\smallskip

Changing the identification of abstract code letters with $q$-ary digits corresponds
to an action of the symmetry group $\Sigma_q$. The main invariants of codes like $k$ and
$d$ only depend on the equivalence class under this action. 

\medskip

{\bf 4.2.2. Proposition.} {\it
The action of the group $\Sigma_q$ induces a unitary equivalence 
of the representations of the Cuntz algebras and a measure preserving 
homeomorphism of the limit sets.}

\smallskip

{\bf Proof.} Suppose given an element $\gamma\in \Sigma_q$ and let $C'=\gamma(C)$ be
the equivalent code obtained from $C$ by the action of $\gamma$. 
The element $\gamma$ induces a map $\gamma: S_C \to S_{C'}$ by 
$$ 
x=x_1 x_2 \cdots x_k \cdots \mapsto \gamma(x)=\gamma(x_1)\gamma(x_2)\cdots \gamma(x_k)
\cdots . 
$$
This map is a homeomorphism. In fact, it is a bijection since $\gamma: C \to C'$ is a bijection,
and it is continuous since the preimage of a clopen set $S_{C'}(w')$ of all words in $S_{C'}$
starting with a given finite word $w'$ consists of the clopen set $S_C(w)$ with $w=\gamma^{-1}(w')$.
Since both $S_C$ and $S_{C'}$ are compact and Hausdorff, the map is a homeomorphism.
It is measure preserving since the measure of the sets $S_C(w)$ is uniform in the words
$w$ of fixed length,
$$ \mu(S_C(w)) =q^{-kr} , \ \ \ \text{ for all } \ \ \  w=w_1,\ldots,w_r, \ \ w_i \in C, $$
so the measure is preserved in permutations of coordinates. 

\smallskip

Thus, the action of $\gamma : S_C \to S_{C'}$ determines a unitary equivalence 
$U_\gamma :L^2(S_{C'}, \mu) \to L^2(S_C, \mu)$, and a representation of the 
algebra $O_C$ on $L^2(S_{C'},\mu)$ generated by  the operators $S_a'=
U_\gamma^* S_a U_\gamma$. This completes the proof.

\medskip

\smallskip
We have seen that, more abstractly, we can identify $\hat S_C$ with the spectrum
of the maximal abelian subalgebra of the algebra $O_C$ generated by the
range projections $S_w S_w^*$, for words $w$ of finite length. One can see in
this way directly that the action of $\Sigma_q$ induces homeomorphisms of
these sets. The uniform distribution of the measure implies that these are
measure preserving.

\medskip

{\bf 4.3. Perron--Frobenius and Ruelle operators.}
Consider again the shift map $\sigma : S_C \to S_C$ defined in (4.2).
The Perron--Frobenius operator $\Cal{P}_\sigma$ is the adjoint of composition by $\sigma$, 
namely
$$
\langle h\circ \sigma, f \rangle = \langle h , \Cal{P}_\sigma f \rangle.
\eqno(4.9)
$$

{\bf 4.3.1. Lemma.}  {\it
The Perron--Frobenius operator $\Cal{P}_\sigma$ is of the form}
$$
\Cal{P}_\sigma = q^{-k/2}   \sum_{a\in C} S_a^*.
\eqno(4.10)
$$

\smallskip

{\bf Proof.} We have
$$ \int_{S_C} \overline{h\circ\sigma} \, \cdot \, f \, d\mu = 
\sum_a \int_{R_a}  \overline{h\circ\sigma} \, \cdot \, f \, d\mu = 
\sum_a \int_{S_C} \overline{h} \, \cdot \, f\circ\sigma_a \, \cdot \, \Phi_a \, d\mu, $$
with $R_a = \sigma_a(S_C)$, so that we have
$$ 
\Cal{P}_\sigma f = \sum_a  \Phi_a f \circ \sigma_a = \sum_a \Phi_a^{1/2} S_a^* f =
q^{-k/2} \sum_a S_a^* f. 
$$
This gives (4.10) and completes the proof.

\medskip

{\bf Remark.} A modified version of the Perron--Frobenius operator which is also useful to
consider is the Ruelle transfer operator for the shift map $\sigma: S_C\to S_C$
with a potential function $W: S_C \to \C$. One usually assumes that the potential
takes non-negative real values. The Ruelle transfer operator $\Cal{R}_{\sigma,W}$ 
is then defined as 
$$
\Cal{R}_{\sigma,W} f (x) = \sum_{y:\sigma(y)=x} W(y)\, f(y).
\eqno(4.11)
$$

\smallskip

For a real valued potential, the operator $\Cal{R}_{\sigma,W}$ is also
obtained as the adjoint of $h \mapsto q^k\,W \cdot h\circ \sigma$, 
$$ \langle q^k\,W\cdot h \circ \sigma, f \rangle = \langle h, \Cal{R}_{\sigma,W} f \rangle, $$
hence it can be regarded as a generalization of the Perron--Frobenius operator. 
The Ruelle and Perron--Frobenius operators are related to the existence of invariant 
measures on $S_C$ and of KMS states for the algebra $O_C$, with respect to
time evolutions related to the potential $W$.

\medskip

{\bf 4.4. Time evolution and KMS states.}
We recall some well known facts about KMS states on the
Cuntz algebras, see for instance [KiKu], [KuRe].

\smallskip
Given a set of real numbers $\{ \lambda_1,\ldots, \lambda_N \}$ there is a time
evolution on the Cuntz algebra $O_N$ which is completely determined by setting
$$
\sigma_t(S_k)= e^{it\lambda_k} S_k.
\eqno(4.12)
$$

Recall that a KMS state at inverse temperature $\beta$ 
on a $C^*$-algebra $\Cal{B}$ with a time evolution $\sigma_t$ is a state 
$\varphi:\Cal{B} \to \C$, such that for each $a,b\in \Cal{B}$ there exists a holomorphic
function $F_{ab}$ on the strip $0< \Im(z)< \beta$, which extends continuously
to the boundary of the strip and satisfies
$$ F_{ab}(t)=\varphi (a\sigma_t(b)), \ \ \ \text{ and } \ \ \  
F_{ab}(t+i\beta)=\varphi(\sigma_t(a)b). $$

\medskip

{\bf 4.4.1. Proposition.} {\it
For the time evolution (4.12) on the Cuntz algebra $O_N$, there exists a unique
KMS state at inverse temperature $\beta>0$ if and only if $\beta$ satisfies}
$$
\sum_{k=1}^N e^{-\beta \lambda_k} =1.
\eqno(4.13)
$$

\smallskip

{\bf Proof.} If $\{ \lambda_1,\ldots, \lambda_N \}$ and a $\beta$ satisfy
(4.13),  then the $\lambda_k$ are all positive and define
$\beta$ uniquely.

As in [KuRe], one uses the Ruelle transfer operator on the
set $X$ of infinite sequences in an alphabet on $N$-letters. 
For a potential $W(x) = e^{-\beta \lambda_{x_1}}$, where $x=x_1x_2\cdots x_n \cdots$,
one finds that the constant function $1$ is a fixed point of $ \Cal{R}_{\sigma,W}$,
$$
 \Cal{R}_{\sigma,W} 1 = (\sum_k e^{-\beta \lambda_k}) 1, 
 $$
hence dually there is a probability measure $\mu_{\lambda,\beta}$ 
on $X$ which is fixed by the dual operator, $\Cal{R}_{\sigma,W}^* \mu_{\lambda,\beta} = \mu_{\lambda,\beta}$.
This is a measure satisfying a self-similarity condition on $X$. In fact, one has
$$
 \Cal{R}_{\sigma,W}^* \mu_{\lambda,\beta} = W \frac{d\mu_{\lambda,\beta} \circ \sigma}{d\mu_{\lambda,\beta}} 
\mu_{\lambda,\beta} ,
 $$
so that $\Cal{R}_{\sigma,W}^* \mu_{\lambda,\beta} = \mu_{\lambda,\beta}$ implies that
$$ 
\frac{d\mu_{\lambda,\beta} \circ \sigma_k}{d\mu_{\lambda,\beta}} = e^{-\lambda_k \beta}, 
$$
and hence $\mu_{\lambda,\beta}$ satisfies the self-similarity condition 
$$ 
\mu_{\lambda,\beta} = \sum_{k=1}^N e^{-\lambda_k \beta} \mu_{\lambda,\beta} \circ \sigma_k^{-1}. 
$$
The measure $\mu_{\lambda,\beta}$ is determined by the values 
$\mu_{\lambda,\beta}(R_k)=e^{-\beta \lambda_k}$, since then the value on a clopen set 
$X(w)\subset X$ of all infinite works starting with a given finite word $w$ of length $r$ is
given by
$$
 \mu_{\lambda,\beta} (X(w))= \int_X  \frac{d\mu_{\lambda,\beta} \circ \sigma^\ell}{d\mu_{\lambda,\beta}} d\mu_{\lambda,\beta}
= e^{-\lambda_{w_1}\beta}\cdots e^{-\lambda_{w_r}\beta}, 
$$
which is consistent with $\mu_{\lambda,\beta} (X(w)) = \sum_{k=1}^N \mu_{\lambda,\beta} (X(wk))$.

\smallskip

By the spectral theory of the operator $\Cal{R}_{\sigma,W}$ one knows, see [KuRe], that
the fixed points  $ \Cal{R}_{\sigma,W} 1 =1$ and $\Cal{R}_{\sigma,W}^* \mu_{\lambda,\beta} = \mu_{\lambda,\beta}$
are unique. This gives then a unique KMS state on $O_N$ at inverse temperature the unique 
$\beta$ satisfying (4.13), which is given by integration with respect to the measure 
$\mu_{\lambda,\beta}$ composed with a continuous linear projection $\Phi: O_N \to C(X)$.

\smallskip

The latter is defined as follows: 
$ \Phi(S_w S_{w'}^*) = 0,$ if $w\ne w^{\prime}$, and  $\chi_{X(w)}$ otherwise,
where $w$ and $w'$ are finite words in the alphabet on $N$ letters.
The state 
$$
\varphi_\beta(S_w S_{w'}^*)=\int \Phi(S_w S_{w'}^*)  d\mu_{\lambda,\beta} = \delta_{w,w'}
e^{-\beta \lambda_{w_1}}\cdots e^{-\beta \lambda_{w_r}},
\eqno(4.14)
$$
for $w$ of length $r$, 
is a KMS state on $O_N$ at inverse temperature $\beta$. One sees that 
it satisfies the KMS condition since it suffices to see that $\varphi_\beta(S_w S_w^*)=
\varphi_\beta(\sigma_{i\beta}(S_{w'}^*) S_w)$. It suffices then to check the latter identity for a single
generator, and use the relations in the algebra to obtain the general case. One has
$\varphi(S_k S_k^*) = e^{-\beta \lambda_k}=\varphi_\beta(e^{-\beta \lambda_k} S_k^* S_k)=
\varphi_\beta(\sigma_{i\beta}(S_k^*) S_k)$.
\smallskip

This completes the proof.

\medskip

{\bf Remark.} Notice that (4.13) can be interpreted as 
the equation that computes the Hausdorff dimension
of a self-similar set where the recursive construction replaces at the first step a set of
measure one with $N$ copies of itself, each scaled by a factor $e^{-\lambda_k}$ and
then iterates the procedure.

In particular, in the main example we are considering here, of the Sierpinski fractal $S_C
\subset Q^n$, the Hausdorff measure $\mu_s$ on $S_C$ with parameter $s=\dim_H(S_C)=k/n$ 
is a self-similar measure as above, and it corresponds to the unique KMS state on the algebra 
$O_C$ at inverse temperature $\beta=\dim_H(S_C)=k/n$, for the time evolution
$$
 \sigma_t (S_a) = q^{-itn} S_a, 
\eqno(4.15)
$$
for all $a \in C$. In fact, in this case the measure satisfies $\mu_s (R_a) = q^{-ns} = q^{-k}$
for all $a\in C$. Thus, the KMS state $\varphi_{k/n}$ takes values
$\varphi_{k/n}(S_w S_w^*)= q^{-k r}$ for a word $w=w_1\cdots w_r$, with 
$w_i=(a_{i1},\ldots,a_{in})\in C$.

\medskip

{\bf 4.5. KMS states and dual traces.} 
Let $\Pi_\ell$ be the set of translates of $\ell$-dimensional intersections of $n-\ell$ 
coordinate hyperplanes. To each $\pi \in \Pi_\ell$ we 
associate a projection in the algebra $O_C$, by taking
$$
P_\pi = \sum_{a\in C\cap \pi} S_a S_a^*. 
\eqno(4.16)
$$
The value of the unique KMS state of $O_C$ at this projection is
$$
 \varphi_{k/n}(P_\pi)=q^{-k} \cdot \#(C\cap \pi)=q^{\ell s -k}, 
\eqno(4.17)
$$
where $s=\dim_H (S_C \cap \pi)$. 

Consider then the algebra obtained by compressing $O_C$ with the projection $P_\pi$,
that is, the algebra generated by the elements $S_{\pi(a)}:=P_\pi S_a P_\pi$. 
These are non trivial when $a\in C\cap \pi$, in which case $S_{\pi(a)}=S_a$, and
zero otherwise, and they satisfy the relations $S_{\pi(a)}^* S_{\pi(a)}=1$, when $S_{\pi(a)}$
is non-trivial, and 
$$ \sum_a  S_{\pi(a)} S_{\pi(a)}^* = P_\pi. $$
Thus, the algebra obtained by compressing with the projection $P_\pi$ is a Toeplitz
algebra $TO_{C\cap \pi}$.

\smallskip

The induced action on the Hilbert space
$L^2(S_C\cap\pi, \mu_s)$ of the algebra $TO_{C\cap \pi}$ obtained as above
descends to the quotient as a representation of $O_{C\cap \pi}$.

\smallskip

On the algebra $O_{C\cap \pi}$ generated by the $S_a$ with $a\in C\cap \pi$,
one can similarly consider a time evolution of the form (4.12), with the 
$\lambda_a$ given by
$$
\lambda_a = - \log \mu_s(R_a),
\eqno(4.18)
$$
where $\mu_s$ is the Hausdorff measure in dimension $s=\dim_H (S_C \cap \pi)$. 
Then one has a unique KMS state on $O_{C\cap \pi}$ at inverse temperature
$\beta = \dim_H (S_C \cap \pi)$, which is determined by integration in this 
Hausdorff measure.

\medskip

In the following we look for a reinterpretation of the Hausdorff dimensions 
considered above in terms of von Neumann dimensions. To this purpose,
we need to consider a type II von Neumann algebra. As we will see below,
there are two ways to associate a type II algebra to the type III algebras $O_C$
that we considered above. The first is passing to the dual system
by taking the crossed product by the time evolution and the second is
considering the fixed point algebra in the weak closure of the GNS representation. 
We finish this subsection by showing that
the first method may not give the needed projections due to the projectionless nature of the resulting 
algebra. We then consider the second possibility in the next subsection, and see 
that one can obtain in that way the desired interpretation as von
Neumann dimensions.

\medskip

It is well known from [Co2] that, 
to a $C^*$-algebra $\Cal{B}$ with time evolution $\sigma_t$, one can associate a dual
system $(\hat \Cal{B},\theta)$, where $\hat\Cal{B} =\Cal{B} \rtimes_\sigma \R$ endowed with a
dual scaling action of $\R^*_+$ of the form 
$\theta_\lambda (\int_\R a(t) U_t dt)=\int_\R \lambda^{it} a(t) U_t dt$.  A KMS state
$\varphi_\beta$ at inverse temperature $\beta$ on $(\Cal{B},\sigma)$ determines a 
dual trace $\tau_\beta$ on $\hat\Cal{B}$, with the scaling condition
$$
\tau_\beta \circ \theta_\lambda = \lambda^{-\beta} \tau_\beta.
\eqno(4.19)
$$
The dual algebra $\hat\Cal{B}$ is generated by elements of the form $\rho(f) a$, with $a\in \Cal{B}$ 
and $f\in L^1(\R)$ and with $\rho(f)=\int_\R f(t) U_t \, dt$. The dual trace is then of the form
$$ \tau_\beta(\rho(f)a)=\varphi_\beta(a) \int_\R \hat f(s) e^{-\beta s} ds, $$
where $\hat f$ is the Fourier transform of $f\in L^1(\R)$. Equivalently, for elements of
the form $f\in L^1(\R,\Cal{B})$ one has $\tau_\beta(f)=\int_\R \varphi_\beta(\hat f(s)) e^{-\beta s} ds$.

\smallskip

If the trace $\tau_\beta$ dual to a KMS state $\varphi_\beta$
is a faithful trace, then, as observed in \cite{Co}, p.586, any projection $P$ in 
$\hat \Cal{A}$ is homotopic to $\theta_1(P)$ so that one should have 
$\tau_\beta(\theta_1(P))=\tau_\beta(P)$, but the scaling property (4.19)
implies that this is also $\tau_\beta(\theta_1(P))=\lambda^{-\beta}\tau_\beta(P)$ 
so that one has $\tau(P)=0$, which by faithfulness gives $P=0$.

\smallskip

{\bf 4.6. Hausdorff dimensions and von Neumann dimensions.}
We show that one can express the Hausdorff dimensions
of the sets $S_C \cap \pi$ in terms of von Neumann dimensions
of projections associated to the linear spaces $\pi$ in the
hyperfinite type II$_1$ factor.

\medskip

{\bf 4.6.1. Proposition.} {\it
Let $C \subset A^n$ be a code with $\# C= q^k$ and let 
$\pi \in \Pi_\ell$ be an $\ell$-dimensional linear space
as above, to which we associate the set $S_C \cap \pi$. To these
data one can associate a projection $P_\pi$ in 
the hyperfinite type II$_1$ factor with von Neumann trace $\tau$, so that 
the von Neumann dimension $\r{Dim}\,(\pi):= \tau(P_\pi)$ is related
to the Hausdorff dimension of $S_C \cap \pi$ by}
$$
\dim_H(S_C\cap \pi)=\frac{k +\log_q \r{Dim}(\pi)}{\ell} ,
\eqno(4.20a)
$$
$$
\dim_H(S_C\cap S_\pi)=\frac{k +\log_q \r{Dim}(\pi)}{n}.
\eqno(4.20b)
$$

\smallskip

{\bf Proof.}
When we consider as above the algebra $O_C$ with the time evolution
$\sigma_t$ of (4.15), we can consider the spectral subspaces of the
time evolution, namely
$$
\Cal{F}_\lambda  =\{ X \in O_C \,|\, \sigma_t(X) =\lambda X \}.
\eqno(4.21)
$$
In particular, $\Cal{F}_0\subset O_C$ is the fixed point subalgebra of the time
evolution. This is generated linearly by elements of the form $S_w S_{w'}^*$, 
for words $w=w_1\cdots w_r$ and $w'=w'_1\cdots w'_r$ word of equal length in elements 
$w_j, w'_j \in C$. The fixed point algebra $\Cal{F}_0$ contains the subalgebra 
$\Cal{A}(\hat S_C)$ identified with the algebra generated by the $S_w S_w^*$.
One has a conditional expectation $\Phi: O_C \to \Cal{F}_0$ given by
$$
\Phi(X) = \int_0^{2\pi/n\log q} \sigma_t(X) dt
\eqno(4.22)
$$
and the KMS state $\varphi_{k/n}$ on $O_C$ is given by
$\varphi_{k/n}=\tau \circ \Phi$, where $\tau$ is the unique normalized
trace on $\Cal{F}_0$, which satisfies
$$ \tau (S_w S_{w'}^*) = \delta_{w,w'} q^{-rk}, $$
for $w$ and $w'$ words of length $r$. This agrees with the values of
the KMS state we saw in (4.14) for $\beta=k/n$ and all the $\lambda_i=n$. 
Consider then the GNS representation 
$\pi_\varphi$ associated to the KMS state $\varphi$ on
$O_C$. We denote by $\Cal{M}$ the von Neumann algebra 
$$
\Cal{M} = \pi_\varphi( O_C)^{\prime\prime}.
\eqno(4.23)
$$
  
By rescaling the time evolution (4.15), the state $\varphi$
becomes a KMS state at inverse temperature
$\beta =1$ for the time evolution 
$$
\alpha_t(S_a) =q^{itk} S_a.
\eqno(4.24)
$$
In fact, we have
$$ \varphi(S_a S_a^*)=q^{-k} = \varphi(\alpha_i(S_a^*) S_a). $$
Thus, up to inner automorphisms,
$\alpha_t$ is the modular automorphism group for the von Neumann algebra
$\Cal{M}$, which shows that the algebra $\Cal{M}$ is of type III$_{q^{-k}}$.
The fixed point subalgebra $\Cal{M}_0$ for the time evolution $\alpha_t$ is the weak
closure of $\Cal{F}_0$. This gives a copy of the hyperfinite type II$_1$ factor $\Cal{M}_0$
inside $\Cal{M}$, with the restriction to $\Cal{M}_0$ of the KMS state $\varphi$ giving the
von Neumann trace $\tau$.  
\smallskip
We then consider the projection $P_\pi=\sum_{a\in C\cap \pi} S_a S_a^*$ as
an element in $\Cal{M}_0$. We have seen that the value of the KMS state $\varphi$
on $P_\pi$ is 
$$ 
\varphi(P_\pi)=\tau(P_\pi)= q^{-k}\cdot  \#(C\cap \pi) = q^{-k + \ell \dim_H(S_C\cap \pi)}
=q^{-k + n \dim_H(S_C\cap S_\pi)}, 
$$
which gives (4.20a) and (4.20b).

\medskip

{\bf 4.7. KMS states and phase transitions for a single code.}
As above, let $C\subset A^n$ be an $[n,k,d]_q$ code and let
$TO_C$ and $O_C$ be the associated Toeplitz and Cuntz algebras,
respectively with generators $T_a$ and $S_a$, for $a\in C$, 
satisfying $T_a^* T_a=1$ for $TO_C$, and $S_a^* S_a=1$
and $\sum_a S_a S_a^* =1$ for $O_C$.

\smallskip
In addition to the representations of $O_C$ on $L^2(S_C,\mu_R)$ constructed previously,
it is natural also to consider the Fock space representation of $TO_C$ on the Hilbert
space $\Cal{H}_C=\ell^2(\Cal{W}_C)$, where $\Cal{W}_C$ is the set of all words of finite length in
the elements $a\in C$, 
$$ 
\Cal{W}_C = \cup_{m \geq 0} \Cal{W}_{C,m}, 
$$
with
$$
 \Cal{W}_{C,m}=\{ w= w_1 \cdots w_m \,|\, w_i \in C \subset  A^n \} 
 $$
and $\Cal{W}_{C,0}:=\{ \emptyset \}$. For all $w$, we identify the words $w\emptyset=w$.
We denote by $\epsilon_w$, for $w\in \Cal{W}_C$, the canonical orthonormal
basis of $\ell^2(\Cal{W}_C)$.  We also denote $\epsilon_\emptyset =\epsilon_0$.

\medskip

{\bf 4.7.1. Lemma.} {\it
The operators on $\Cal{H}_C$ given by
$$
T_a \epsilon_w = \epsilon_{aw}
\eqno(4.25)
$$
define a representation of the Toeplitz algebra 
$TO_C$ on $\Cal{H}_C$.}

\smallskip

{\bf Proof.} The adjoint $T_a^*$ of the operator (4.25) 
is given by
$$
T_a^* \epsilon_w = \delta_{a,w_1} \epsilon_{\sigma(w)},
\eqno(4.26)
$$
where $\delta_{a,w_1}$ is the Kronecker delta, and $\sigma(w)=w_2\cdots w_m \in \Cal{W}_{C,m-1}$, 
for $w=w_1\cdots w_m \in \Cal{W}_{C,m}$. In fact, we have
$$
 \langle T_a f, h\rangle = \sum_w \overline{f_{aw}} h_w  =
\sum_{w'=aw} \overline{f_{w'}} h_{\sigma(w')}  =
\sum_{w'} \overline{f_{w'}} \delta_{a,w_1'}  h_{\sigma(w')} =  \langle f, T_a^* h\rangle, 
$$
for $f=\sum_w f_w \epsilon_w$ and $h=\sum_w h_w \epsilon_w$ in $\Cal{H}_C$.
Thus, $T_a T_a^*=P_a$, where $P_a$ is the projection onto the subspace $\Cal{H}_{C,a}$
spanned by the $\epsilon_w$ with $w_1 =a$. One also has
$$
 T_a^* T_a f = \sum_w f_w T_a^* \epsilon_{aw} = f, 
 $$
so that we obtain $T_a^* T_a=1$.

\smallskip

This completes the proof.

\medskip

We consider then time evolutions on the algebra $TO_C$ associated to
the random walks and Ruelle transfer operators introduced in \S 4.3 and 4.4. 

\medskip

{\bf 4.7.2. Lemma.} {\it
Let $W_\beta(x)=\exp(-\beta \lambda_{x_1})$, for $x\in S_C$, be a potential satisfying
the Keane condition $\sum_{a\in C} \exp(-\beta \lambda_a) =1$. 
Then setting
$$
\sigma_t(T_a) = e^{it \lambda_a} T_a 
\eqno(4.27)
$$
defines a time evolution on the algebra $TO_C$, which is implemented, in the
Fock representation, by the Hamiltonian}
$$
H  \epsilon_w  = (\lambda_{w_1}+\cdots +\lambda_{w_m}) \, \, \epsilon_w, \ \ \ \text{ for } w=w_1\cdots w_m \in \Cal{W}_{C,m}.
\eqno(4.28)
$$

\smallskip 

{\bf Proof.} It is clear that (4.27) determines a 1--parameter group of
continuous automorphisms of the algebra $TO_C$.
The Hamiltonian that implements the time evolution in the Fock representation
is a self adjoint unbounded operator on the Hilbert space $\Cal{H}_C$
with the property that $\sigma_t(A)= e^{it H}  A e^{-it H}$,  for all elements
$A\in TO_C$. We see on the generators $T_a$ that
$$ e^{it H}  T_a e^{-it H} \epsilon_w = e^{it (\lambda_a + \lambda_{w_1}+\dots +\lambda_{w_n})}
e^{-it( \lambda_{w_1}+\dots + \lambda_{w_n})} \epsilon_{aw}  $$
implies that $e^{itH}$, with $H$ as in (4.28) is the one-parameter group
that implements the time evolution (4.27) in the Fock representation.

\smallskip

The proof is completed.

\smallskip

We consider in particular the time evolution associated to the uniform Hausdorff
measure on the fractal $S_C$ of dimension $R=k/n$.

\medskip

{\bf 4.7.3. Proposition.} {\it 
For an $[n,k,d]_q$- code $C$, we consider the time evolution
$$ \sigma_t(T_a) = q^{it n} T_a  $$
on the algebra $TO_C$.  Then for all $\beta >0$ there is a unique KMS$_\beta$
state on the resulting quantum statistical mechanical system. 
\smallskip
(1) At low temperature 
$\beta > R$, this is a type I$_\infty$ state, with 
the partition function given by $Z_C(\beta)=\r{Tr}(e^{-\beta H})= 
(1-q^{(R-\beta)n})^{-1}$ and the Gibbs equilibrium state of the form
$$
 \varphi_\beta(A)=Z_C(\beta)^{-1}\,\, \r{Tr}(A e^{-\beta H}) . 
\eqno(4.29)
$$

(2) At the critical temperature $\beta =R$, the unique KMS$_\beta$ state
is a type III$_{q^{-k}}$ factor state, which induces the unique KMS
state on the Cuntz algebra $O_C$, and is determined by 
the normalized $R$-dimensional Hausdorff measure $\mu_R$ on $S_C$.
It is given by the residue
$$
\varphi_R (A) = \r{Res}_{\beta=R} \r{Tr}(A e^{-\beta H}).
\eqno(4.30)
$$

(3) At high temperature the unique KMS state is also of type III and determined 
by the values $\varphi_\beta(T_w T_w^*)= e^{-\beta (\lambda_{w_1}+\cdots +\lambda_{w_m})}$,
where $\lambda_a = n \log q$ for all $a\in C$.
\smallskip
(4) Only at the critical temperature $\beta=R$ the KMS state $\varphi_R$ 
induces a KMS state on the quotient algebra $O_C$.}

\smallskip

{\bf Proof.}
First notice that any KMS state at inverse temperature $\beta$ must
have the same values on elements of the form $T_w T_{w'}^*$. This can
be seen from the KMS condition, inductively from 
$$ \varphi_\beta (T_a T_a^*) = \varphi_\beta (\sigma_{iR}(T_a^*) T_a) = q^{-R n} 
\varphi_\beta (T_a^* T_a)= q^{-\beta n}. $$
This determines the state uniquely. So we see that at all $\beta >0$ where
the set of KMS states is non-empty it consists of a single element.

\smallskip

The Hamiltonian has eigenvalues $m n \log q$, for $m\in \N$, each 
with multiplicity $q^{m k_r} = \# \Cal{W}_{C,m}$. Thus, the partition function of
the time evolution is given by
$$
Z_C(\beta)=\r{Tr}(e^{-\beta H}) =
$$
$$
= \sum_m \sum_{w\in \Cal{W}_{C,m}} \exp(-\beta (\lambda_{w_1}+\cdots +\lambda_{w_m})) =
$$
 $$
=\sum_m q^{m k} q^{-\beta n m} =\sum_m q^{(R-\beta)nm} . 
\eqno(4.31)
$$
The series converges for inverse temperature
$\beta > R$, with sum $$Z_C(\beta)=(1-q^{(R-\beta)n})^{-1}.$$
Thus, in the low temperature range $\beta > R$, one has an equilibrium state of
the Gibbs form (4.29).

\smallskip

At the critical temperature $\beta =R$, we have a KMS$_\beta$ state
of type III$_{q^{-k}}$, which is the unique KMS state on the algebra $O_C$
$$
\varphi_R (A)=\int_{S_C} \Phi(A)\, d\mu_R,
\eqno(4.32)
$$
which induces a KMS state on $TO_C$ by pre--composing the 
expectation $\Phi: O_C \to \Cal{A}(\hat S_C)$ with the quotient map
$TO_C \to O_C$. Here we use again the identification of $\Cal{A}(\hat S_C)$
with the maximal abelian subalgebra of $O_C$, and $\mu_R$ is
the normalized $R$-dimensional Hausdorff measure on $S_C$.
This means that the state $\varphi_R$ has values
$$ \varphi_R(T_w T_{w'}^*)= \delta_{w,w'} \mu_R (S_C(w))= q^{-R n m} = q^{-k m}, $$ 
for $w=w_1\ldots w_m$. To see that, at this critical temperature, the state
is given by a residue (and can therefore be expressed in terms of Dixmier trace),
it suffices to observe that the partition function $Z(\beta)$ has a simple pole at
$\beta =R$ with residue $\r{Res}_{\beta =R} Z(\beta)=1$, so that we have
$$ 
\r{Res}_{\beta =R} \r{Tr}(T_w T_{w'}^* \, e^{-\beta H}) =
e^{-\beta (\lambda_{w_1}+\cdots + \lambda_{w_m})}  \r{Res}_{\beta =R}  Z(\beta) =
\varphi_R(T_w T_{w'}^*). 
$$
At higher temperatures $\beta < R$ the KMS state is similarly determined by the list
of values 
$$ 
\varphi_R(T_w T_{w'}^*)= \delta_{w,w'} e^{-\beta (\lambda_{w_1}+\cdots + \lambda_{w_m})}
= \delta_{w,w'} q^{-\beta n m}. 
$$
\smallskip

To see that only the state at critical temperature induces a KMS state on the
quotient algebra $O_C$ it suffices to notice that in $O_C$ one has the
additional relation $\sum_a S_a S_a^* =1$, which requires that the values
of a KMS$_\beta$ state satisfy the Keane relation
$$ \sum_a \varphi_\beta(S_a S_a^*) = \sum_a e^{-\beta \lambda_a} =1 . $$
This is satisfies at $\beta=R$, where it gives the self-similarity relation for
the Hausdorff dimension of the fractal $S_C$, but it is not satisfied at any other
$\beta\neq R$. 
\smallskip

The proof is complete.

\smallskip

We see from the above result that the situation is very similar to the one
encountered in the construction of the Bost--Connes system [BoCo], 
where the case of the system without interaction is obtained as a
tensor product of Toeplitz algebras (in that case in a single generator)
with their unique KMS$_\beta$ state at each $\beta >0$. We explain below
how a similar approach with tensor products plays a role here in describing
the curve $R=\alpha_q(\delta)$ in terms of phase transitions.

\medskip

{\bf 4.8. Crossed product description.}
Before we discuss families of codes and tensor products of quantum
statistical mechanical systems, it is worth reformulating the setting
described above in a way that may make it easier to pass to 
the analog of the ``systems with interaction" of [BoCo].

\smallskip

Let $C$ be an $[n,k,d]_q$ code.
We introduce the notation $\Xi_C(P)$ for the algebra obtained by compressing the
abelian subalgebra $T\Cal{A}(C) \subset TO_C$ with a projection $P$ of $TO_C$,
$$ \Xi_C(P) := P\,\, T\Cal{A}(C)\,\, P. $$

\smallskip

The isometries $T_a$, for $a\in C$, determine an endomorphism $\rho$ of
the algebra $T\Cal{A}(C)$ given by
$$
\rho(X) = \sum_a T_a \, X \, T_a^* .
\eqno(4.33)
$$
This endomorphism satisfies $\rho(1) =P$, the idempotent 
$\sum_a T_a T_a^* =P$ in $T\Cal{A}(C)\subset TO_C$. 
The endomorphism
$\rho$ has partial inverses $\sigma_a$ given by
$$
\sigma_a(X) = T_a^* X T_a ,
\eqno(4.34)
$$
for $X\in \Xi_C(P_a)$, where $P_a=T_a T_a^*$ is the range projection.
They satisfy
$$
\sigma_a \rho(X) = X, \ \ \ \forall X\in T\Cal{A}(C).
\eqno(4.35)
$$

Notice that, for $X=T_w T_w^*$ in $T\Cal{A}(C)$, we have
$P X = T_{w_1} T_{w_1}^* T_w T_w^* = T_w T_w^*=X$
and $X P = T_w T_w^* T_{w_1} T_{w_1}^* = T_w T_w^*=X$,
so that, if one represents an arbitrary element $X\in T\Cal{A}(C)$
in the form $X= \lambda_0 + \sum_w \lambda_w T_w T_w^*$,
one finds $PX = XP= \sum_a \lambda_0 T_a T_a^* +
\sum_w \lambda_w T_w T_w^*$. Similarly, one has
$\rho(X)= \lambda_0 P + \sum_{aw} \lambda_w T_a T_w T_w^* T_a^*$,
which acts as a shift on the coefficients $\lambda_w$ and lands in
the compressed algebra $\Xi_C(P)$.
The partial inverses $\sigma_a$ satisfy $\sigma_a(1)=1$
since $T_a^* T_a=1$, and they map an element
$X= \lambda_0 + \sum_w \lambda_w T_w T_w^*$ of $T\Cal{A}(C)$
to $\sigma_a(X)= \lambda_0 + \sum_{w=aw'} \lambda_w T_{w'} T_{w'}^*$.

\smallskip

In the case of the quotient algebra $O_C$, where one imposes the relations
$S_a^*S_a=1$ and $\sum_a S_a S_a^*=1$, the endomorphism above induces an
endomorphism $\bar\rho$ of the algebra $\Cal{A}(\hat S_C)$ with $\bar\rho(1)=1$, which
is given simply by the composition 
$$ \bar\rho(f)= \sum_a S_a \, f \, S_a^* = f\circ \sigma $$
with the one-sided  shift map $\sigma: S_C \to S_C$,
$$ \sigma(x_1x_2\cdots x_m \cdots) = x_2 x_3 \cdots x_{m+1} \cdots $$
and the partial inverses are the compositions with the partial inverses of
the one sides shift
$$ \bar\sigma_a(f) = S_a^* \, f \, S_a = f\circ \sigma_a, $$
where $\sigma_a(x_1x_2\cdots x_m \cdots)=ax_1x_2\cdots x_m \cdots$.

\smallskip

Thus, we can form the crossed product algebra $T\Cal{A}(C)\rtimes_{\rho} \bold{M}$,
where $\bold{M}$ is the additive monoid $\bold{M}=\Z$.
This has generators $T_w T_w^*$ together with an extra generator $S$ satisfying
$S^* S=1$ and $S X S^* = \rho(X)$. 
It also satisfies $SS^*=P$ and $S^* X S =\sigma_a(X)$, for $X\in \Xi_C(P_a)$.

\medskip

{\bf 4.8.1. Proposition.} {\it
The morphism $\Psi: TO_C \to T\Cal{A}(C)\rtimes_\rho \bold{M}$ defined by setting
$$
\Psi(T_a) = P_a S 
\eqno(4.36)
$$
identifies $TO_C$ with the subalgebra $\Xi_C(P)  \rtimes_\rho \bold{M}$.
On the quotient algebra $O_C$, the induced morphism $\bar\Psi$ gives an isomorphism 
$O_C \simeq \Cal{A}(\hat S_C)\rtimes_\rho \bold{M}$. }

\smallskip

{\bf Proof.} Notice
that $\Psi(T_a)^* \Psi(T_a)= S^* P_a S=\sigma_a(P_a)=T_a^* P_a T_a=1$
and $\sum_a \Psi(T_a) \Psi(T_a)^*=\sum_a P_a S S^* P_a=\sum_a P_a P P_a=\sum_a P_a=P$,
since, as observed above, $P_a P =P P_a =P_a$. Thus, $\Psi$ maps injectively
$TO_C\subset T\Cal{A}(C)\rtimes_\rho \bold{M}$. To see that surjectivity also holds, notice that 
$\Xi_C(P) \rtimes_\rho \bold{M}$ is spanned linearly by monomials of the form $T_w T_w^* S^k$
and $S^k T_w T_w^*$, for $w\in \Cal{W}_{C,m}$, $m\geq 1$,
and $k\geq 0$. It suffices to show that these are
all in the range of the map $\Psi$. First observe that the map $\Psi$ is the identity on the
subalgebra $T\Cal{A}(C)\subset TO_C$. In fact, for $w=w_1\cdots w_m$, with $w_i\in C$, we have
$$ \Psi(T_w T_w^*)= P_{w_1}\rho(P_{w_2})\cdots \rho^{m-1}(P_{w_m}) (S S^*)^m
 \rho^{m-1}(P_{w_m}) \cdots P_{w_1} $$ $$ = P_w P P_w= P_w = S_w S_w^*. $$
Notice then that we have $\Psi(\sum_a T_a) = \sum_a P_a S = PS$.  Let $Y=\sum_a T_a$
in $TO_C$. We then have
$$ \Psi(T_w T_w^*) \Psi(Y^k)= T_w T_w^* (PS)^k. $$
We have $(PS)^k = P\dots \rho^{k-1}(P) S^k$. Since $P =S S^*$ and $\rho(X)=S X S^*$,
we see that $P \rho(P)=\rho(P)$ and $P\rho(P)\cdots \rho^{k-1}(P)=\rho^{k-1}(P)=S^{k-1} {S^*}^{k-1}$.
Thus, $\rho^{k-1}(P) S^k = S^k$ and we obtain that 
$$ \Psi(T_w T_w^* Y^k) = T_w T_w^* S^k. $$
The argument for elements of the form $S^k T_w T_w^*$ is analogous. Thus, all the monomials
with $w\in \Cal{W}_{C,m}$ with $m\geq 1$ are in the range of $\Psi$ and the only missing terms
are the $S^k$ and their adjoints (the case of $w=\emptyset \in \Cal{W}_{C,0}$). 

\smallskip

This induces the isomorphism $O_C\simeq \Cal{A}(\hat S_C)\rtimes_{\bar\rho}\bold{M}$
of \cite{Exel}, where in the quotient algebra $\bar S^* f \bar S=q^{-k} \sum_a f\circ \sigma_a$
is the Perron--Frobenius operator and the induced map $\bar\Psi$ preserves the 
additional relation $\sum_a S_a S_a^*=1$. Thus, in this case we have $\bar\Psi(\sum_a S_a)=
\sum_a P_a \bar S= \bar S$, since in this case $\bar P=\sum_a S_a S_a^*= 1$. 
We then obtain that the range of $\bar\Psi$ is all of $\Cal{A}(\hat S_C)\rtimes_{\bar\rho}\bold{M}$.
This completes the proof.
\smallskip
With this description of the algebra $TO_C$ in terms of crossed product of $\Xi_C(P)$
by the monoid $\bold{M}$, one can view the time evolution as given by
$$
\sigma_t (X) = X, \ \ \ \text{ for } \ \ X \in \Xi_C(P), \ \ \ \text{ and } \ \ \ \sigma_t(S)= q^{it n} S.
\eqno(4.37)
$$



\newpage

\bigskip

\centerline{\bf 5. Quantum statistical mechanics and Kolmogorov complexity}

\medskip

Our reformulation of the rate and relative minimum distance of codes
in terms of Hausdorff dimensions, as well as the construction of algebras
with time evolutions for individual codes, can be reinterpreted within the
context of Kolmogorov complexity and Levin's universal enumerable semi-measures.

\medskip
{\bf 5.1. Languages and fractals.}
We begin with some considerations on structure functions and entropies for codes.
Suppose given a code $C \subset A^n$, for an alphabet $A$ with
$\# A =q$. We assume that $C$ is an $[n,k,d]_q$ code.

First we reinterpret the construction of the fractal $S_C$ in terms of
languages and $\omega$-languages. 

Given the alphabet $A$, one writes $A^\infty=\cup_n A^n$ for the set of all 
words of finite length in the alphabet $A$ and one denotes by $A^\omega$ 
the set of all words of infinite length in the same alphabet. A language $\Lambda$
is a subset of $A^\infty$ and an $\omega$-language is a subset of $A^\omega$.

To a code $C$ one can associate a language $\Lambda_C$ given by
all words in $A^\infty$ that are successions of words in $C\subset A^n$.
Similarly, one has an $\omega$-language $\Lambda_C^\omega$ given
by all infinite words in $A^\omega$ that are a succession of elements in $C$.
As such, the $\omega$-language $\Lambda_C^\omega$ is set-theoretically
identified with the fractal $\hat S_C$ we considered previously.  

There is a notion of entropy for languages ([Eilen], see also the recent [Sta3]), 
which is defined as follows. One first introduces the structure function 
$$
s_\Lambda(m) = \# \{ w \in \Lambda\, : \, \ell(w) =m \},
$$
the number of words of length $m$ in the language $\Lambda$. These can be assembled
together into a generating function 
$$
G_\Lambda(t) = \sum_m s_\Lambda(m) t^m.
$$
The entropy of the language $\Lambda$ is then the log of the radius of 
convergence of the series above
$$
\Cal{S}_\Lambda = - \log_{\# A} \rho(G_\Lambda). 
$$

{\bf 5.1.1. Lemma.} {\it
For the language $\Lambda_C$ defined by an $[n,k,d]_q$-code $C$ 
the structure function satisfies
$$ G_{\Lambda_C}(q^{-\beta}) = Z_C(\beta), $$
where $Z_C(\beta)$ is the partition function of the quantum statistical mechanical
system $(TO_C,\sigma_t)$ associated to the code $C$. The entropy of the language
$\Lambda_C$ is the rate of the code $\Cal{S}_{\Lambda_C}=k/n=R$.}

\smallskip

{\bf Proof.}
In the case of an $[n,k,d]_q$-code $C$, notice that the series $G_{\Lambda_C}$ is
given by
$$
G_{\Lambda_C}(t)= \sum_m q^{km} t^{nm} = (1-q^k t^n)^{-1} ,
$$
since one has  $s_\Lambda(N)= 0$ for $N\neq m n$, while for $N=m n$ one has
$s_\Lambda(nm)=q^{km}$.
In particular, when expressed in the variable $t = q^{-s}$ this becomes
$$
G_{\Lambda_C}(q^{-s})= \sum_m q^{(R-s)nm} = (1-q^{(R-s)n})^{-1} ,
$$
with convergence for $\beta=\Re(s) > R$. This recovers the partition function
$Z_C(\beta)$ of the quantum statistical mechanical system associated to the
code $C$. This gives an entropy 
$$ \Cal{S}_{\Lambda_C}=- \log_q \rho(G_\Lambda) = R=k/n, $$
since domain of convergence for $\beta > R$ corresponds to $|t| =|q^{-s}| < q^{-R}$.

\smallskip

Intersection with linear spaces $\pi_\ell$ determines induced languages
$\Lambda_{C,\ell}$. The threshold value $\ell=d$ corresponds to the 
minimal dimension for there is a choice of $\pi_d$ for which the resulting
language is non-trivial, with entropy $d$.

\medskip

{\bf 5.2. Kolmogorov complexity.}
There are several variants of Kolmogorov complexity for words $w$ of finite length in a
given alphabet, see [LiVi], \S 5.5.4. To any such complexity function $K(w)$
one associates the {\rm lower Kolmogorov complexity} for infinite words
by setting
$$
\kappa(x) = \liminf_{w\to x} \frac{K(w)}{\ell(w)},
$$
where the limit is taken over finite words $w$ that are truncations of increasing
length $\ell(w)=m\to \infty$ of an infinite word $x$.
There is a characterization (see [ZvoLe] and [LiVi]) of the lower Kolmogorov complexity
in terms of measures, which we discuss more at length in the case of codes here below.

\smallskip

We begin by recalling the notion of semi-measures and provide examples taken
from the constructions we have already seen in the previous sections of this paper.

\medskip

{\bf 5.2.1. Definition.} {\it A semi-measure on $S_C$ is a positive real valued function
on the cylinder sets $\{ S_C(w) \}$ that satisfies $\mu(S_C)\leq 1$ and the 
subadditivity property
$$ \mu(S_C(w)) \geq \sum_{a\in C} \mu(S_C (wa)). $$
}

Here we do not distinguish between $\hat S_C=\Lambda_C^\omega$ and 
$S_C$ since the difference is of measure zero in any of the above measures.
An example of semi-measures is obtained using the Ruelle transfer operator
techniques considered above.

\medskip

{\bf 5.2.2. Lemma.} {\it
Let $W_\beta(x)$ be a potential that satisfies the Keane condition at $\beta=\beta_0$
and such that, for a fixed $x$, it is monotonically decreasing as a function of $\beta$.
Then the function 
$$
\mu_{x_0,\beta}(S_C(w))=W_\beta(w_1 x_0)\cdots W_\beta(w_n \cdots w_1 x_0)
$$
is a semi-measure.}

\smallskip

{\bf Proof.}
Suppose given a potential $W_\beta(x)$, and assume that for a $\beta=\beta_0$
it satisfies the Keane condition $\sum_{a\in C} W_{\beta_0}(ax)=1$. Assume,
moreover, that for fixed $x\in S_C$, the function $W_\beta(x)$ is monotonically 
decreasing as a function of $\beta$. This will certainly be the case for the special 
cases we considered with $W_\beta(x)=e^{-\beta \lambda_{x_1}}$ of 
$W_\beta(x)=e^{-\beta \lambda_{x_1 x_2}}$. One will then have
$$ \sum_{a\in C} W_\beta(ax) \leq 1, \ \ \ \text{ for } \ \ \beta \geq \beta_0, \ \ \ \forall x\in 
S_C . $$
Thus, one has
$$ \sum_{a\in C} \mu(S_C (wa)) =\sum_{a\in C}
W_\beta(w_1 x_0)\cdots W_\beta(w_n \cdots w_1 x_0) \cdot W_\beta(a w_n \cdots w_1 x_0) $$
$$ \leq W_\beta(w_1 x_0)\cdots W_\beta(w_n \cdots w_1 x_0) =
\mu_{x_0,\beta}(S_C(w)) , $$
for all $\beta \geq \beta_0$. This completes the proof.

\medskip

{\bf 5.3. Enumerable semi-measures.}
In complexity theory one is especially interested in those semi-measures that are
enumerable. We recall here a characterization of enumerable semi-measure given in
Theorem 4.5.2 of [LiVi], which will be useful in the following,

Given a language $\Lambda$, let $\Cal{F}_\Lambda$ be the class of functions 
(called monotone in [LiVi])  $f: A^\infty \to \Lambda$, where $A^\infty$ 
is the set of all finite words (of arbitrary length) in the alphabet $A$,
with $f(w w')=f(w) f(w')$, the product being concatenation of words in 
$\Lambda$. These extend to functions from $A^\omega$, the set of
all infinite words in the alphabet $A$ to the $\omega$-language
$\Lambda^\omega$.

Given a semi-measure $\mu$ on $A^\omega$ and a function $f\in \Cal{F}_\Lambda$
one obtains a semi-measure $\mu_f$ on $\Lambda^\omega$ by setting
$$ \mu_f (\Lambda^\omega(w)) = \sum_{w'\in A^\infty: f(w')=w} \mu(A^\omega(w')), $$
where, as usual, $\Lambda^\omega(w)$ and $A^\omega(w')$ denote the subsets
of $\Lambda^\omega$ and $A^\omega$, respectively, made of infinite words starting 
with the given prefix word $w$ or, respectively, $w'$.

In particular, let $\lambda$ denote the 1-dimensional Lebesgue measure on $[0,1]$.
This induces a measure on $A^\omega$ by mapping the infinite sequences in $A^\omega$
to points of $[0,1]$ written in their $q$-ary expansion. The measure satisfies
$$ \lambda (A^\omega(w)) = q^{-\ell(w)}, $$
where $\ell(w)$ is the length of the word $w\in A^\infty$.

Then Theorem 4.5.2 of [LiVi] characterizes enumerable semi-measures
on $\Lambda^\omega$ as those semi-measures $\mu$ that are obtained
as $\mu =\lambda_f$ for a function $f\in \Cal{F}_\Lambda$.

\smallskip

We observe first that these measures satisfy the following multiplicative property. 
For simplicity of notation, we write in the following $\mu(w)$ for $\mu(\Lambda^\omega(w))$.

\medskip

{\bf 5.3.1. Lemma.} {\it The enumerable semi-measures are multiplicative on concatenations of words, 
$\mu(ww')=\mu(w) \mu(w')$.}

\medskip

{\bf Proof.} The uniform Lebesgue measure $\lambda$ clearly has that property since
$\lambda(ww')= q^{-\ell(ww')}=q^{-(\ell(w)+\ell(w'))}=\lambda(w)\lambda(w')$. Suppose
then given a function $f\in \Cal{F}_\Lambda$. This satisfies $f(ww')=f(w)f(w')$ by definition.
Thus, in particular, we can write $f(w)=f(w_1)\cdots f(w_m)$, for a word $w=w_1\cdots w_m$
of length $\ell(w)=m$. Consider then the measure $\mu=\lambda_f$ given by
$\lambda_f (u) = \sum_{w: f(w)=u} \lambda(w)$. For a word $u=u_1 \cdots u_m$ of
length $\ell(u)=m$, we can then write this equivalently as
$$ \lambda_f(u) = \sum_{f(w_i)=u_i} \prod_i \lambda(w_i) =
\prod_{i=1}^m \lambda_f(u_i). $$
This completes the proof.

\smallskip

The characterization of enumerative semi-measure as semi-measures of the form 
$\mu=\lambda_f$ shows, for example, that the uniform Hausdorff measure of dimension
$\dim_H S_C=R=k/n$ on the set $S_C$ considered above is an enumerative (semi)-measure.
In fact, it is of the form $\mu = \lambda_f$, where the map $f$ is induced by the coding map
$E: A^k \to C \subset A^n$, so that elements $a\in C$ are described as $a=f(w)$ for a
word $w\in A^k$. In this case, since the coding map $E$ is injective, there is a unique
word $w$ with $f(w)=a$. 

\smallskip

Another example of an enumerative (semi)-measure on $S_C$ can be obtained using as function
$f \in \Cal{F}_{\Lambda}$ the decoding map $P$, by which we mean the map that assigns to each 
element in $A^n$ the nearest point in $C$ in the Hamming metric. Then one obtains
$$ \mu_f (S_C(u)) = \sum_{w=(w_i): w_i \in A^n \, , \,  P(w_i)=u_i} \lambda(w) =
\#\{ w=(w_i):  P(w_i)=u_i \} q^{-n m},  $$ 
for $u =u_1 \cdots u_m$ with $u_i \in C$, and $w=w_1 \cdots w_m$ with $w_i \in A^n$.

\smallskip

We now connect enumerable semi-measures on $S_C$ to quantum statistical
mechanical systems on the Toeplitz--Cuntz algebra $TO_C$ in the following way.

\medskip

{\bf 5.3.2. Lemma.} {\it Let $\mu$ be a semi-measure on $S_C$ such that $\mu(ww')=\mu(w)\mu(w')$,
where $\mu(w)$ is shorthand for $\mu(S_C(w))$. Then setting
$$ \sigma_t(T_a) = \mu(S_C(a))^{-it} T_a $$
determines a time evolution $\sigma_t \in Aut(TO_C)$. In the Fock space representation 
of $TO_C$, this time evolution has Hamiltonian   
$$ H \epsilon_w = - \log \mu(S_C(w))\, \epsilon_w . $$ 
The partition function is
$$ Z_{\mu,C}(\beta)= (1-\sum_{a\in C} \mu(S_C(a))^\beta )^{-1}, $$
with a pole at a critical $\beta_c\leq 1$, the inverse
temperature at which $\sum_a \mu(a)^{\beta_c}=1$. 
The functional 
$$ \varphi(T_w T_{w'}^*)= \delta_{w,w'} \, \mu(S_C(w))^\beta $$
is a KMS$_\beta$ state for the quantum statistical mechanical system $(TO_C, \sigma)$.
}

\medskip

{\bf Proof.} In the Fock representation the time evolution is generated by a Hamiltonian
$$ e^{itH} T_a e^{-itH} \epsilon_w = \sigma_t(T_a) \epsilon_w =\mu(a)^{-it} \epsilon_{aw} , $$
which gives
$$  e^{itH} \epsilon_w = \mu(w)^{-it} \epsilon_w $$
using the fact that the semi-measure satisfies $\mu(aw)=\mu(a)\mu(w)$. This gives
$H \epsilon_w = -\log \mu(w)$. The partition function is then given by
$$ Z_{\mu,C}(\beta) =Tr(e^{-\beta H})= \sum_w \mu(w)^{\beta}. $$
Again using $\mu(w)=\mu(w_1)\cdots \mu(w_m)$ for $w=w_1\cdots w_m$ a
word of length $\ell(w)=m$, we write the above as
$$ \sum_w \mu(w)^{\beta}= \sum_m \sum_{w_1,\ldots,w_m} \mu(w_1)^\beta \cdots
\mu(w_m)^\beta = \sum_m (\sum_{a\in C} \mu(a)^\beta)^m. $$
For $\beta > \beta_c$ where $\sum_a \mu(a)^{\beta_c}=1$, the series converges to
$$ Z_{\mu,C}(\beta)=(1-\sum_{a\in C} \mu(a)^\beta )^{-1}. $$
Since $\mu$ is a semi-measure, it satisfies $\sum_a \mu(a) \leq 1$, so that $\beta_c\leq 1$. 
The state defined by the condition $\varphi(T_w T_{w'}^*)=\delta_{w,w'} \mu(w)^\beta$ 
satisfies the KMS$_\beta$ condition. This can be checked inductively from
$$ \varphi(T_a T_a^*)= \mu(a)^\beta =\mu(a)^\beta \varphi(T_a^* T_a) =\varphi
(T_a^* \sigma_{i\beta}(T_a)). $$
This completes the proof.

\medskip

This result in particular shows that, given a semi-measure on $S_C$ with strict inequality
$\sum_a \mu(a) <1$, there is a way to associate to it a measure by raising
the temperature, that is, lowering $\beta$ from $\beta=1$ to $\beta =\beta_c$.
One then has $\varphi(S_w S_w^*)=\mu(w)^{\beta_c}$, this
time satisfying the correct normalization $\sum_a \mu(a)^{\beta_c}=1$,
which also implies
$$ \sum_a \mu(wa)^{\beta_c} =\mu(w)^{\beta_c} \sum_a \mu(a)^{\beta_c} 
= \mu(w)^{\beta_c}, $$
so that one indeed obtains a measure. 

\medskip

{\bf 5.4. Universal enumerable semi-measure.}
A well known result of Levin (see [ZvoLe] or Theorem 4.5.1 of [LiVi]) is that there exist
universal (or maximal)  enumerable semi-measures $\mu_U$ on $\Lambda^\omega$. They are
characterized by the following property: any enumerable semi-measure $\mu$ is absolutely 
continuous with respect to $\mu_U$ with bounded Radon-Nikodym derivative, or 
equivalently $\mu_U \geq c_f \lambda_f $,  for all $f\in \Cal{F}_\Lambda$. Such universal
semi-measures are not unique. A way to construct one is by listing the enumerable semi-measures
(or equivalently listing the functions $f\in \Cal{F}_\Lambda$) and then taking
$\mu_U= \sum_n \alpha_n \lambda_{f_n}$ with positive real coefficients $\alpha_n$ with
$\sum_n \alpha_n \leq  1$, see Theorem 4.5.1 of [LiVi]. Another description which is more
suitable for our purposes is as an enumerable semi-measure $\mu_U =\lambda_{f_U}$,
where $f_U$ is a {\it universal monotone machine} in the sense of Definitions 4.5.2 and 4.5.6
of [LiVi], that is, universal for Turing machines with a one-way read-only input tape, some
work tapes, and a one-way write-only output tape. As an enumerable semi-measure, we
can apply to it the construction of a corresponding time evolution and quantum statistical 
mechanical system as above. Notice that $\mu_U$ is not recursive and it is not a measure,
that is, the inequality $\sum_a \mu_U(a) <1$ is strict, see Lemma 4.5.3 of [LiVi].

\medskip

We can then consider on the Toeplitz--Cuntz algebra $TO_C$ the {\it universal time evolution}
$$ \sigma_t (T_a) = \mu_U(a)^{-it} T_a $$
induced by the universal enumerable semi-measure $\mu_U=\lambda_{f_U}$. The critical
value $\beta_U <1$ at which the partition function
$$ Z_{U,C}(\beta)= (1 - \sum_a \mu_U(a)^\beta)^{-1} $$
has a pole is the {\it universal critical inverse temperature}.  This universal critical temperature
can be regarded as another parameter of a code $C$, which in this setting replaces the
code rate $R$ as the critical $\beta$ is the time evolution. 

\smallskip

The universal critical inverse temperature $\beta_U$ can also be described as a 
Hausdorff dimension, by modifying the construction of the Sierpinski fractal $S_C$
associated to the code $C$ in the following way. 

Recall that $S_C$ is constructed inductively starting with the space 
$(0,1)_q^n$ viewed as $(\infty \times n)$-matrices with entries in $A$. 
At the first step, replacing it by $q^k$ copies scaled down by a factor of
$q^{-n}$, each identifies with the subset $(0,1)_{q,a}^n$ of points in $(0,1)_q^n$
where the first row is equal to the element $a\in C$, with $C\subset A^n$.
Each $(0,1)_{q,a}^n$ is a copy of $(0,1)_q^n$ scaled down by a factor of $q^{-n}$.
One obtains then $S_C$ by iterating this process on each $(0,1)_{q,a}^n$ and
so on.

Now we consider a very similar procedure, where we again start with the same set
$(0,1)_q^n$. We again consider all the subsets $(0,1)_{q,a}^n$ as above, but where
the set $(0,1)_{q,a}^n$ is metrically a scaled down copy of $(0,1)_q^n$, now scaled
by a factor $\mu_U(a)$ instead of being scaled by the uniform factor $q^{-n}$ as in
the construction of $S_C$. One obtains in this way a fractal $S_{C,U}$, by
iterating this process. The self similarity equation for the non-uniform 
fractal $S_{C,U}$ is then given by
$$ \sum_{a\in C} \mu_U(a)^s =1, $$
which identifies its Haudorff dimension with $s=\beta_U$.

\smallskip
 
One also has a Ruelle transfer operator associated to the universal enumerable
semi-measure, which is given by
$$ \Cal{R}_{\sigma, U,\beta} f (x) = \sum_{a\in C} \mu_U(a)^\beta\, f(ax). $$

\smallskip

It is then natural to investigate 
how the universal enumerative semi-measure is related to the Hausdorff dimension 
$\dim_H S_C=R$ and to Kolmogorov complexity.

\medskip

{\bf 5.4.1. Lemma.} {\it For all words $x\in \hat S_C$ the lower Kolmogorov complexity is
bounded above by
$$ \kappa(x) \leq \dim_H(S_C)=R. $$ }

\medskip

{\bf Proof.} The universal enumerable semi-measure $\mu_U$ is related to the lower
Kolmogorov complexity by ([UShe], [ZvoLe], [Sta3])
$$
\kappa(x) = \liminf_{w\to x} \frac{-\log_q \mu_U(w)}{\ell(w)},
$$
where again the limit is taken over finite length truncations $w$ of the infinite word $x$
as the length $\ell(w)$ goes to infinity. We know by construction that the universal $\mu_U$
dominates multiplicatively all the enumerable semi-measures. Thus, in particular, if $\mu$
is the Hausdorff measure on $S_C$ of dimension $R=\dim_H(S_C)$, which we have seen
above is an enumerable (semi)-measure, there is a positive real number $\alpha$ such
that $\mu_U(w)\geq \alpha \mu(w)$, for all finite words $w$. This implies that
$$ \frac{-\log_q \mu_U(w)}{\ell(w)} \leq \frac{ -\log_q \mu(w)}{\ell(w)} + \frac{ - \log_q \alpha}{\ell(w)}. $$
This gives
$$  \liminf_{w\to x}  \frac{-\log_q \mu_U(w)}{\ell(w)} \leq \lim_{w\to x} \frac{ -\log_q \mu(w)}{\ell(w)} = 
\lim_{m\to \infty} \frac{k m}{nm} =R. $$

\medskip

Moreover, we have the following result.

\medskip

{\bf 5.4.2. Lemma.} {\it The lower Kolmogorov complexity satisfies
$$ \sup_{x\in \hat S_C} \kappa(x) = R $$
with the supremum achieved on a set of full measure.}

\medskip

{\bf Proof.} This follows directly from Ryabko's inequality [Rya1], [Rya2], which shows that
in general one has the estimate
$$ \dim_H(\Lambda^\omega) \leq \sup_{x\in \Lambda^\omega} \kappa(x). $$
To see this more explicitly in our case, recall first that the Hausdorff dimension of a
set $X$ embedded in some larger ambient Euclidean space can be computed in 
the following way. Consider coverings $\{ U_\alpha \}$ of $X$ with diameters
$diam(U_\alpha)\leq \rho$ and consider the sum $\sum_\alpha diam(U_\alpha)^s$.
Set $$ \ell_s(X,\rho)=\inf \{ \sum_\alpha diam(U_\alpha)^s\,:\,  diam(U_\alpha)\leq \rho \}. $$
Then one has 
$$ \dim_H(X)= \inf \{ s\,:\, \lim_{\rho\to 0} \ell_s(X,\rho) =0 \} =
\sup \{ s\,:\, \lim_{\rho\to 0} \ell_s(X,\rho) =\infty \}. $$
We then use an argument similar to the one used in [Rya2]: from  
$$ \kappa(x) = \liminf_{w\to x} \frac{-\log_q \mu_U(w)}{\ell(w)} $$
we know that, for a given $x\in S_C$, and for arbitrary $\delta>0$, there is an integer 
$m(x)$ such that, if $w(x)$ denotes the truncation of length $m(x)$ of the infinite word 
$x$ then
$$  \frac{-\log_q \mu_U(w(x))}{m(x)} \leq \kappa(x) + \delta \leq \kappa + \delta, $$
where $\kappa = \sup_x \kappa(x)$ as above. The integer $m(x)$ can be taken so
that $q^{-m(x)} \leq \rho$ for a given size $\rho\in \R^*_+$. Let $\Cal{L}$ be the
countable set of words $w=w(x)$ of lengths $m(x)$, for $x\in S_C$, obtained as above.
We can then construct a covering of $S_C$ with sets $S_C(w)$, for $w\in \Cal{L}$, with
diameters $diam(S_C(w)) = \sqrt{n}\,q^{-m(x)} \leq \sqrt{n}\, \rho$, for a positive constant
$\alpha$ that only depends on $n$. These satisfy
$$ \sum_{w\in \Cal{L}} diam(S_C(w))^{\kappa + \delta} \leq  \alpha
\sum q^{-m(x) (\kappa +\delta)} , $$
with $\alpha=\sqrt{n}^{(\kappa +\delta)}$. This gives
$$ \sum q^{- m(x) (\kappa +\delta)} \leq \sum q^{m(x) \frac{\log_q \mu_U(w(x))}{m(x)} }
\leq \sum_{w\in \Cal{L}} \mu_U(w) \leq 1. $$
We then have
$$ \ell_s (S_C,\rho) \leq \sum_{w\in \Cal{L}} diam(S_C(w))^s  $$
and therefore
$$ \lim_{\rho\to 0} \ell_s (S_C,\rho) \leq \sum_{w\in \Cal{L}} diam(S_C(w))^s. $$
For $s=\kappa +\delta$ the right hand side is uniformly bounded above,
so $\lim_{\rho\to 0} \ell_{\kappa +\delta} (S_C,\rho) < \infty$, hence $\kappa+\delta \geq \dim_H(S_C)$, hence $\kappa \geq \dim_H(S_C)$, since $\delta$ can be chosen 
arbitrarily small.


\bigskip

\centerline{\bf 6. Functional analytic constructions for limit points}

\medskip
{\bf 6.1. Realizing limit points of the code domain.}
We have seen in the previous sections that, given an $[n,k,d]_q$ code $C$,
one can construct fractal sets $S_C$ and $S_\pi$ as in \S 3.3, that have Hausdorff dimension,
respectively, equal to $R=k/n$ and $\delta=d/n$, and that the parameter $d$
can be characterized in terms of the behavior of the Hausdorff dimension
of the intersections $S_{C,\ell,\pi}=S_C \cap S_\pi$ for $\pi$ of dimension $\ell$.
We now consider the case where two assigned values $R$ and $\delta$ 
are not necessarily realized by a code $C$, but are an accumulation point
of the code domain, namely there exists an infinite family $C_r$ of $[n_r,k_r,d_r]_q$
codes, where $k_r/n_r \to R$ and $d_r/n_r \to \delta$ as $r\to \infty$.

\smallskip

We show here that one can still construct sets $S_R$ and $S_\delta$,
depending on the approximating family $C_r$, with
the property that $\dim_H(S_R)=R$ and $\dim_H(S_\delta)=\delta$
and so that these sets are, in a suitable sense, approximated by the
sets $S_{C_r}$ and $S_{\pi_r}$ with $\pi_r\in \Pi_{d_r}$ of the family
of codes $C_r$.

\medskip

{\bf 6.2. Multifractals in infinite dimensional cubes.}
Let then $(0,1)_q^\infty$ denote the union $(0,1)_q^\infty=\cup_n (0,1)_q^n$ which can
be considered as direct limit under the
inclusion maps that embed $[0,1]^n\subset [0,1]^{n+1}$ as the face
in $[0,1]^{n+1}$ of which the last coordinate is equal to zero. This is a metric space with the 
induced metric. In terms of the $q$-ary expansion, elements in $(0,1)_q^\infty$ can be 
described as infinite matrices with only finitely many columns with non zero entries.
We can embed all the
$S_{C_r}\subset (0,1)_q^{n_r}$ of an approximating family inside $(0,1)_q^{\infty}$. Thus, we
can view the set $S_R=\cup_r S_{C_r}$ as $S_R\subset (0,1)_q^\infty$. 

\medskip

{\bf 6.2.1. Proposition.} {\it (1)
For any limit point $(R,\delta)$ of the code domain there exists a
family $C_r$ of $[n_r,k_r,d_r]_q$ codes with $k_r/n_r \nearrow R$ and $d_r/n_r \nearrow \delta$. 
(2) For such a sequence $C_r$ 
the sets $S_R=\cup_r S_{C_r}$ and $S_\delta = \cup_r S_{\pi_{d_r}}$ have
$$
\dim_H (S_R)=R,  \ \ \ \  \dim_H(S_\delta) =\delta, \ \ \ \text{ and } \ \ \ 
\dim_H (S_R\cap S_\delta) >0.
\eqno(6.1)
$$
(3) Moreover, given a sequence $\pi_{\ell_r}\in \Pi^{(n_r)}_{\ell_r}$ with $\ell_r \leq d_r-1$,
one can form the analogous $S_\ell = \cup_r S_{\pi_{\ell_r}}$. This has the property that
$\dim_H(S_R\cap S_\ell)=0$.}
 
\smallskip

{\bf Proof.} (1) We first show that we can find an approximating family $C_r$ with
$k_r/n_r \nearrow R$ and $d_r/n_r \nearrow \delta$. To this purpose we
use the spoiling operations on codes described above. We know from
Corollary 1.2.1 that, given an $[n,k,d]_q$ code, we can produce
an $[n,k-1\leq k' \leq k,d-1]_q$ code from it by applying the second and third spoiling
operations and twice the first one. Starting with an approximating family
$C_r$ with $k_r/n_r\to R$ and $d_r/n_r\to \delta$ and using the spoiling operations
as described, we can produce from it other approximating families with $k_r$
replaced by $k_r-\ell_r$ and $d_r-\ell_r$ with $\ell_r/n_r\to 0$ and such that,
for sufficiently large $r$, $k_r/n_r - \ell_r/n_r \leq R$ and $d_r/n_r -\ell_r/n_r \leq \delta$.
Possibly after passing to a subsequence, we obtain a family where the new
$k_r$ and $d_r$ satisfy $k_r/n_r \nearrow R$ and $d_r/n_r \nearrow \delta$. 

\smallskip

(2) 
The Hausdorff dimension of a union behaves like $$\dim_H (\cup_r X_r) = \sup_r 
\dim_H(X_r)$$ by countable stability ([Fal], p.~37). 
Thus, if $k_r/n_r \nearrow R$ and $d_r/n_r \nearrow \delta$, we
obtain that $\dim_H(S_R)=R$ and $\dim_H(S_\delta)=\delta$.

\smallskip
Let us now show that $\dim_H(S_R\cap S_\delta)>0$. We have  
$S_R\cap S_\delta = \cup_r (S_{C_r}\cap S_{\pi_d})$. Again by countable stability of
the Hausdorff dimension we obtain
$$ \dim_H(S_R\cap S_\delta) = \sup_r \dim_H(S_{C_r}\cap S_{\pi_d}) > 0. $$
The Hausdorff dimension is also bounded above by the dimension of $S_R$ and $S_\delta$
so $0<\dim_H(S_R\cap S_\delta) \leq \min\{ R,\delta \}$.

(3) For a given sequence $\ell_r\leq d_r-1$ with corresponding linear spaces
$\pi_{\ell_r}\in \Pi^{(n_r)}_{\ell_r}$, we can form the sets
$S_{\pi_{\ell_r}} \subset (0,1)_q^\infty$.
If the $\ell_r$ are chosen so that the ratio sequence $\ell_r /n_r \nearrow \ell$ approaches a limit
from below as $r\to \infty$, then the same argument given above shows that the
Hausdorff dimension $\dim_H(\cup_r S_{\pi_{\ell_r}})=\ell$. For $S_\ell =\cup_r S_{\pi_{\ell_r}}$, 
the intersection $S_R\cap S_\ell$ is given as above by
$S_R\cap S_\ell = \cup_r (S_{C_r}\cap S_{\pi_{\ell_r}})$. Since $\ell_r\leq d_r -1$,
we know that $\dim_H(S_{C_r}\cap S_{\pi_{\ell_r}})=0$ for all $r$. Thus, we have
$\dim_H(S_R\cap S_\ell) =0$. This shows that the set $S_\delta$ still has the 
same threshold property with respect to the behavior of the Hausdorff dimension
of the intersection with $S_R$, as in the case of the individual $S_C$ of a single code.

\medskip
{\bf 6.3. Random processes and fractal measures for limit points of codes.}
We have seen how, for an individual code $C\subset A^n$ we can construct a
fractal set $S_C$ of Hausdorff dimension the code rate $R$ and with the Hausdorff
measure $\mu_R$ in dimension $R$ satisfying the self-similarity condition
$$ \mu_R = q^{-nR} \sum_{a\in A^n} \mu_R \circ \sigma_a^{-1} . $$
\smallskip
We now consider the case of a limit point $(R,\delta)$, which is an accumulation point 
of the code domain, so that we have a family of codes $C_r$ with $k_r/n_r\to R$
and $d_r/n_n\to \delta$. As we have seen in Proposition 6.2.1 above, 
we can construct a set $S_R \subset (0,1)_q^\infty$ with Hausdorff dimension $\dim_H(S_R)=R$.

The construction of $S_R$ shows that the Hausdorff dimension of each
$S_{C_r}$ is dominated by that of the larger ones and of $S_R$. Therefore
for the uniform $R$-dimensional Hausdorff measure each of the $S_{C_r}$ becomes 
negligible. However, it is possible to construct non-uniform measures on $S_R$ 
that give non-trivial probability to each of the $S_{C_r}$.
We investigate here how to obtain self-similar multifractal measures on the sets $S_R$
using the method of Ruelle transfer operators.

\smallskip

On the set $S_R\subset (0,1)_q^\infty$ we consider a  potential $W=W_\beta$
with non-negative real values satisfying the Keane condition
$$
\sum_a W_\beta (ax) =1, \ \ \ \ \forall x\in S_R,
\eqno(6.2)
$$
where for $x\in S_{C_r} \subset S_R$ the sum is over all the elements
$a\in C_r$.

The Ruelle transfer operators on $S_R$ will then be of the form
$$
\Cal{R}_{\sigma,W} f(x) = \sum_{\sigma(y)=x} W(y) f(y)= \sum_{a\in \cup_r (C_r\cap A^{n_r})} W(ax) f(ax),
\eqno(6.3)
$$
where the shift map $\sigma$ on $S_R$ is the one induced by the shift maps on the
individual $S_{C_r}$. The partial inverses of $\sigma$ are given by
maps $\sigma_a(x)=ax$, where, for $x\in S_{C_r}$, $a$ is an element of corresponding $C_r$.

\smallskip

{\bf Example 1.} One can consider the case where the potential $W_\beta(x)$ 
is a piecewise constant function on $S_R$, which depends only on the first coordinate (first row)
$x_1\in \cup_r(C_r\cap A^{n_r})$ of $x$. One can write it in this case as
$$
W_\beta(x) = e^{-\beta \lambda_{x_1}}, \ \ \ \text{ with } \ \ \  \sum_{a\in \cup_r (C_r\cap A^{n_r})} 
e^{-\beta \lambda_a} =1.
\eqno(6.4)
$$

{\bf Example 2.} Another case we will consider in the following is where the potential is also
a piecewise constant function on $S_R$, but which depends on the
first two coordinates (first two rows) $x_1,x_2\in \cup_r (C_r\cap A^{n_r})$ 
of $x\in (0,1)_q^\infty$. In this case we
write it in the form 
$$
W_\beta(x) = e^{-\beta \lambda_{x_1x_2}}, \  \  \  \text{ with } \  \  \  \sum_{a\in \cup_r (C_r\cap A^{n_r})} 
e^{-\beta \lambda_{ax_1}} =1, 
\eqno(6.5)
$$
for all $x_1 \in \cup_r (C_r\cap A^{n_r})$.
We then think of $\lambda_{ab}$ as an infinite matrix indexed by elements $a,b\in \cup_r (C_r\cap A^{n_r})$. 
The condition that $\sum_a W_\beta(ax)=1$ for all $x\in S_R$ implies that the function
$f(x)=1$ is a fixed point for the transfer operator $\Cal{R}_{\sigma,W,\beta}$. 

\smallskip

Here is a version of the construction given in [DutJor] (see also for instance [MarPa]), 
for an arbitrary potential $W_\beta$ satisfying the Keane condition.

\medskip

{\bf 6.3.1. Proposition.} {\it
For a choice of a point $x_0\in S_R$, 
one can then construct a measure $\mu_{\beta x_0}$ on $S_R$ 
by assigning to the subset $S_R(w)\subset S_R$  of words
$x\in S_R$ that start with a given finite length word $w=w_1\cdots w_m$ 
with $w_j\in \cup_r (C_r\cap A^{n_r})$ the measure}
$$
\mu_{\beta,x_0} (S_R(w)) = W_\beta(w_1 x_0) W_\beta(w_2 w_1 x_0)\dots 
W_\beta(w_n \dots w_1 x_0).
\eqno(6.6)
$$

\smallskip
 
{\bf Proof.}
To see that this indeed defines a probability measure we need to check that
$$ 
\sum_w \mu_{\beta,x_0}(S_R(w)) =1, 
$$
and that
$$
 \sum_{a\in \cup_n A^n} \mu_{\beta,x_0}(S_R(wa)) = \mu_{\beta,x_0}(S_R(w)). 
 $$
The first condition is satisfied since we have
$$
 \sum_{w_1\cdots w_n} W_\beta(w_1 x_0) W_\beta(w_2 w_1 x_0)\cdots 
W_\beta(w_{n-1} \cdots w_1 x_0)
W_\beta(w_n \cdots w_1 x_0) = 
$$  
$$ 
\sum_{w_1\cdots w_{n-1}}W_\beta(w_1 x_0) W_\beta(w_2 w_1 x_0)\cdots  W_\beta(w_{n-1} \cdots w_1 x_0) =\dots  $$ $$ =\sum_{w_1}  W_\beta(w_1 x_0) =1, 
$$
by repeatedly using the Keane condition (6.2). The second condition also follows from
(6.2), since we have
$$ 
\sum_a \mu_{\beta,x_0}(S_R(wa)) =
\sum_a W_\beta(w_1 x_0) \cdots 
W_\beta(w_n \cdots w_1 x_0)
W_\beta(a w_n \cdots w_1 x_0)
$$ 
$$ = W_\beta(w_1 x_0) W_\beta(w_2 w_1 x_0)\cdots 
W_\beta(w_n \cdots w_1 x_0), 
$$
since $\sum_a W_\beta(a w_n \cdots w_1 x_0) =1$.
\smallskip
This completes the proof.

\medskip

The idea is that one thinks of the measure constructed as above as the probability of a random walk that starts at $x_0$ and proceeds at each step in the direction marked by an element $a\in \cup_r (C_r\cap A^{n_r})$.
In the special cases (6.4) and (6.5), the probabilities are given, respectively, by
$$ \mu_{\beta,x_0} (S_R(w)) = \prod_{j=1}^m e^{-\beta \lambda_{w_j}}, $$
which is, in this case, independent of the choice of the point $x_0$, and by
$$ \mu_{\beta,x_0} (S_R(w)) = e^{-\beta \lambda_{w_n w_{n-1}}}  \cdots
e^{-\beta \lambda_{w_2 w_1}} e^{-\beta \lambda_{w_1 x_0}}. $$

Consider then a fixed $S_{C_r}$ inside $S_R=\cup_r S_{C_r}$. 
The measure constructed as above on $S_R$
induces a multi-fractal measure on each $S_{C_r}$. We describe the resulting system
of measures explicitly in the two cases where the measure on $S_R$ satisfies
(6.4) or (6.5).

\medskip

{\bf 6.3.2. Proposition.}{\it
(1) If the measure on $S_R$ satisfies (6.4), then it induces on each $S_{C_r}$
a multi--fractal measure by assigning
$$
 \mu_{\beta,r}(S_{C_r}(w))= \frac{1}{Z_r(\beta)^m} \prod_j  e^{-\beta \lambda_{w_j}}, 
\eqno(6.7)
$$
for $w=w_1\cdots w_m$ with $w_i\in C_r$, where $Z_r(\beta)$ is given by
$$
 Z_r(\beta)= \sum_{a \in C_r} e^{-\beta \lambda_a} .
\eqno(6.8)
$$
\smallskip
(2) If the measure on $S_R$ satisfies (6.5), then it induces on each $S_{C_r}$
a multi-fractal measure by assigning
$$
\mu_{\beta,r,x_0}(S_{C_r}(w)) = \frac{
W_\beta(w_m w_{m-1}) \cdots W_\beta (w_1 x_0)\, f^{(r)}_{w_m}}{
 \rho_{\beta,r}^m \, f_{x_0}^{(r)} },
\eqno(6.9)
$$ 
for $w=w_1\cdots w_m$ with $w_i\in C_r$,
where $f^{(r)}$ is the Perron--Frobenius eigenvector of the positive matrix $W_\beta(ab)=e^{-\beta \lambda_{ab}}$ and $\rho_{\beta,r}$ the eigenvalue equal to the spectral radius.}

\smallskip
{\bf Proof.}
When one restricts the potential $W_\beta$ from $S_R$ to a single $S_{C_r}$, the infinite
sum (6.2) is replaced by a truncated finite sum
$$
\sum_{a\in C_r\cap A^{n_r}} W_\beta (ax) < 1, \ \ \ \ \forall x\in S_R.
\eqno(6.10)
$$
Thus, in the case (6.4), instead of the normalization condition given by the infinite sum
$$ 
\sum_{a\in \cup_r C_r} e^{-\beta \lambda_a} =1, 
$$
we have a partition function given by the finite sum (6.8).
The induced probability measure on $S_{C_r}$ is then given by assigning measures
$$
 \mu_{\beta,r}(S_{C_r}(a))= \frac{e^{-\beta \lambda_a}}{Z_r(\beta)}, 
 $$
and more generally by (6.7) on the sets $S_{C_r}(w)$ with
$w=w_1\cdots w_m$ with $w_i\in C_r$. Since $Z_r(\beta)^{-1} \sum_{a\in C_r} e^{-\beta \lambda_a}=1$, this assignment satisfies the required properties in order to define a probability measure on $S_{C_r}$. Notice that the measure obtained in this way is no longer a uniform self-similar measure like the Hausdorff measure on $S_{C_r}$ of Hausdorff dimension $k_r/d_r$, but it is a non-uniform multi-fractal measure in the sense of  [Fal], \S 17.

The case where the potential $W_\beta$ on $S_R$ satisfies (6.5) is similar. The restriction of $W_\beta$ to a single $S_{C_r}$ gives a $q^{k_r}\times q^{k_r}$-matrix, $W_\beta(ab)=e^{-\beta \lambda_{ab}}$, for $a,b\in C_r$. This matrix is positive, in the sense that all its entries are, by construction, positive real numbers. Thus, the Perron--Frobenius theorem applied to the matrix $W_\beta(ab)$ (or rather to its transpose) shows that there exists a unique eigenvector $f^{(r)}=(f^{(r)}_a)$ 
$$
\sum_{a\in C_r} W_\beta(ab) f^{(r)}_a = \rho_{\beta,r} \, f_b^{(r)},
\eqno(6.11)
$$
with positive entries $f^{(r)}_a >0$ and with eigenvalue $\rho_{\beta,r}$ equal to the spectral radius of $W_\beta(ab)$.

\smallskip 
We then show that setting the measure of $S_{C_r}(w)$ equal to (6.9), 
for $w=w_1\cdots w_m$ with $w_j \in C_r$,
defines an induced probability measure on $S_{C_r}$. We check that
$$ 
\sum_w \mu_{\beta,r,x_0}(S_{C_r}(w)) = \sum_{w_1\cdots w_m}
\frac{W_\beta(w_m w_{m-1}) \cdots W_\beta (w_1 x_0)\, f^{(r)}_{w_m}}
{\rho_{\beta,r}^m  \, f_{x_0}^{(r)} } 
$$
$$ = \sum_{w_1\cdots w_{m-1}} 
\frac{W_\beta(w_{m-1} w_{m-2}) \cdots W_\beta (w_1 x_0)\, f^{(r)}_{w_{m-1}} }
{\rho_{\beta,r}^{(m-1)} \,  f_{x_0}^{(r)} } $$
$$ 
= \sum_{w_1}  \frac{W_\beta (w_1 x_0)\, f^{(r)}_{w_1} }{\rho_{\beta,r} \, f_{x_0}^{(r)} }
= 1, 
$$
since we have
$$ 
\sum_{w_{j+1}} W_\beta(w_{j+1} w_j) f^{(r)}_{w_{j+1}} = \rho_{\beta,r} f^{(r)}_{w_j}.
 $$
Similarly, we have
$$  
\sum_a \mu_{\beta,r,x_0}(S_{C_r}(wa)) = \sum_a \frac{W_\beta(a w_m) \dots W_\beta (w_1 x_0)\,  f^{(r)}_a} {\rho_{\beta,r}^{m+1}  \, f_{x_0}^{(r)} } = 
$$
$$
 \frac{W_\beta(w_m w_{m-1}) \cdots W_\beta (w_1 x_0)\, f^{(r)}_{w_m}}
{\rho_{\beta,r}^m  \, f_{x_0}^{(r)} } = \mu_{\beta,r,x_0}(S_{C_r}(w)) ,
$$
since we have
$$
 \sum_a W_\beta(a w_m) f^{(r)}_a = \rho_{\beta,r} f^{(r)}_{w_m} . 
 $$
We therefore obtain a family of induced multi-fractal probability measures on the $S_{C_r}$.
\smallskip

This completes the proof.

\smallskip

A similar construction can be done in the case of the family of sets $S_{\pi_{d_r}}$ with
$d_r/n_r\nearrow \delta$ and the set $S_\delta=\cup_r S_{\pi_{d_r}}$.
\medskip

{\bf 6.4. Limit points and algebra representations.}
As above, consider a family of codes $C_r$ with parameters 
$k_r/n_r\nearrow R$ and $d_r/n_r \nearrow \delta$. We have Toeplitz
algebras $TO_{C_r}$ associated to each code in this family. It is then
natural to consider as algebra associated to the limit point $(R,\delta)$
the infinite Toeplitz algebra in the union of the generators of all the
$TO_{C_r}$, namely $TO_{\cup_r C_r}$ generated by isometries
$S_a$ for $a\in \cup_r C_r$.

\medskip
{\bf 6.4.1. Proposition.} {\it
Let $\mu_{\beta,x_0}$ be a probability measure on $S_R$ constructed as above,
in terms of a potential $W_\beta(x)$. The algebra $TO_{\cup_r C_r}$ has a representation
on the Hilbert space $L^2(S_R,\mu_{\beta,x_0})$ given by
$$
(S_a f)(x) = W_\beta(ax_0)^{-1/2}\, \chi_{S_R(a)}(x)\, f(\sigma(x)), 
\eqno(6.12)
$$
for $a\in \cup_r C_r $.}

\smallskip
{\bf Proof.} We must check that the operators (6.12), for $a\in \cup_r C_r$, 
satisfy the relations $S_a^* S_a=1$ of $TO_{\cup_r C_r}$, with $S_a S_a^*=P_a$ 
orthogonal range projections. 

First observe that the Radon-Nikodym derivative of $\mu_{\beta,x_0}$ with respect to
composition with $\sigma_a$ for $a\in \cup_r C_r$ satisfies
$$
\frac{d\mu_{\beta,x_0}\circ \sigma_a}{d\mu_{\beta,x_0}} = W_\beta(ax_0).
\eqno(6.13)
$$
In fact, we have
$$ \mu_{\beta,x_0}(S_R(w))=\sum_a \mu_{\beta,x_0}(S_R(wa)) = $$ $$
\sum_a \int_{S_R(w)} \frac{d\mu_{\beta,x_0}\circ\sigma_a}{d\mu_{\beta,x_0}} d\mu_{\beta,x_0}
= \sum_a W_\beta(ax_0) \mu_{\beta,x_0}(S_R(w)). $$

It then follows that the operators $S_a$ of (6.12) have adjoints 
$$
(S_a^* f)(x) = W_\beta(ax_0)^{1/2} f(\sigma_a(x)). 
\eqno(6.14)
$$
In fact, we have 
$$ \langle S_a h, f \rangle = \int_{S_R(a)} W_\beta(ax_0)^{-1/2}\, \overline{h(\sigma(x))}\, f(x)\,
 d\mu_{\beta,x_0}(x)  $$
$$ = \int_{S_R} W_\beta(ax_0)^{-1/2}\, \overline{h(u)} \, f(\sigma_a(u)) \, \frac{d\mu_{\beta,x_0}\circ\sigma_a}{d\mu_{\beta,x_0}} d\mu_{\beta,x_0}(u) $$ $$ =\int_{S_R} \overline{h(u)} \, 
W_\beta(ax_0)^{1/2}\,f(\sigma_a(u)) \, d\mu_{\beta,x_0}(u) =\langle h, S_a^* f\rangle. $$
One then sees explicitly that the operators $S_a$ and $S_a^*$ satisfy $S_a^* S_a=1$,
while $S_a S_a^*$ is the range projection $P_a$ given by multiplication by the characteristic
function $\chi_{S_R(a)}$. Notice that, for $a\neq a'$ in $\cup_r C_r$, the sets $S_R(a)$ and
$S_R(a')$ are disjoint, hence the range projections are orthogonal. Thus, we obtain a representation
of the algebra $TO_{\cup_r C_r}$.

\smallskip

This completes the proof.
\smallskip
One can proceed in a similar way with respect to the parameter $\delta$ using the
set $S_\delta$ with a similar measure and representation. Thus, the choice of a limit
point $(R,\delta)$ corresponds to the pair of Hilbert spaces $L^2(S_R,\mu_{\beta,x_0})$
and $L^2(S_\delta, \mu_{\beta',x_0'})$ with representations of the algebras $TO_{\cup_r C_r}$
and $TO_{\cup_r \pi_{d_r}}$, respectively. 

\smallskip

The main asymptotic problem of codes (\cite{Man}, \cite{TsfaVla}) consists of identifying a
continuous curve $R=\alpha_q(\delta)$ (which can also be symmetrically formulated
as $\delta=\alpha'_q(R)$) that gives for fixed $\delta$ the maximal possible value of $R$
in the closure of the subset of limit points of the code domain 
(respectively, the maximal $\delta$ for fixed $R$).
We describe here a way to characterize the curve $R=\alpha_q(\delta)$ in terms of
the measures $\mu_{\beta,x_0}$ on the sets $S_R$ and the uniform self-similar measures
on the $S_{C_r}$ for approximating families of codes.

\smallskip
We have shown earlier that given a point $(R,\delta)$ in
the closure of the code domain, it is always possible to construct an approximating
family of codes $C_r$ with $k_r/n_r\nearrow R$ and $d_r/n_r \nearrow \delta$. In
the following, we refer to such a family $\{ C_r \}$ as a {\it good approximating family}.
\smallskip
We have shown that a measure $\mu_{\beta,x_0}$ on the set $S_R\subset (0,1)_q^\infty$
induces a compatible family of non-uniform fractal measures on the sets $S_{C_r}\subset (0,1)_q^{n_r}$.
We now show that, conversely, the family of uniform self-similar measures on the
$S_{C_r}$ determine a family of
non-uniform measure $\mu_{\beta,x_0}$ on the set $S_R\subset (0,1)_q^\infty$, for
$\beta > R$.

\medskip

{\bf 6.5. Proposition.} {\it
Let $C_r$ be a good approximating family for a limit point $(R,\delta)$.
For $a\in \cup_r C_r$ set $\lambda_a = n_r \log q$, where $n_r$ corresponds to 
the smallest $C_r\subset (0,1)_q^{n_r}$ for which $a\in C_r$. Then the series 
$$
 Z_{\cup_r C_r}(\beta) := \sum_{a\in \cup_r C_r} e^{-\beta \lambda_a} 
\eqno(6.15)
$$
converges for $\beta > R$ and the potential
$$
 W_\beta(x) = Z_{\cup_r C_r}(\beta)^{-1} \, \exp(-\beta \lambda_{x_1}) 
\eqno(6.16)
$$
defines a probability measure on the set $S_R$. The analogous construction
holds for $S_\delta$ with convergence in the domain $\beta > \delta$.}

\smallskip
{\bf Proof.} We have
$$ 
Z_{\cup_r C_r}(\beta) = \sum_r q^{k_r} q^{-\beta n_r}, 
$$
since the $S_{C_r}$ are disjoint in $(0,1)_q^\infty$.
Since $\{ C_r \}$ is a good approximating family, we have $k_r/n_r \leq R$ and we see
that 
$$ 
\sum_r q^{k_r} q^{-\beta n_r} \leq \sum_r q^{(R-\beta)n_r}. 
$$
This is convergent for $\beta > R$. The potential $W_\beta(x)$ of (6.16)
then satsifies the Keane condition $\sum_a W_\beta(ax)=1$. The construction for
$S_\delta$ is entirely analogous, using the uniform measures on the $S_{\pi_{d_r}}$.
This completes the proof.

\smallskip
We then obtain the following characterization of the curve $R=\alpha_q(\beta)$
of the fundamental asymptotic problem for codes.

\medskip

{\bf 6.6. Proposition.} {\it
The domain $\beta \geq \alpha_q(\delta)$ is the closure of the common 
domain of convergence of the functions $Z_{\cup_r C_r}(\beta)$ for all
the points $(R,\delta)$ with fixed $\delta$ in the closure of the subset of limit points 
of the code domain and for all good approximating families $\{ C_r \}$.}

\smallskip

{\bf Proof.}
The domain $\beta \geq  R$ is in fact the closure of the common domain of convergence of the
functions $Z_{\cup_r C_r}(\beta)$ when one varies the good approximating family $C_r$.
In fact, the argument above shows that they all converge for $\beta > R$. The $S_{C_r}$ are
disjoint in $(0,1)_q^\infty$ so that the zeta function (6.15) is given by 
$\sum_r q^{k_r} q^{-\beta n_r}$. Then if $\beta < R$, for sufficiently large $r$ one will
have $k_r/n_r -\beta > 0$ and the series diverges. 
Then by varying the limit point $(R,\delta)$ with fixed $\delta$ one obtains the result.

\medskip

{\bf Remark.} We constructed in \S 6.4 multi-fractal measures on the set $\cup_r S_{C_r}$
for a family of codes $\{ C_r \}$ approximating a limit point $(R,\delta)$. We also considered,
associated to the same family of codes, the infinite Toeplitz algebra $TO_{\cup_r C_r}$. Notice
that in this case, unlike what happens for the case of a single code, the set $\cup_r S_{C_r}$ 
is no longer dense in the spectrum of the maximal abelian subalgebra. In fact, the latter 
consists of all infinite sequences in the elements of $\cup_r C_r$, while the set
$\cup_r S_{C_r}$ only contains those sequences where all the successive elements
in an infinite sequence belong to the same $C_r$. Both sets can be regarded as
the union of the $\omega$-languages defined by the codes $C_r$, where in the case
of $\cup_r S_{C_r}$ one is keeping track of the information of the embeddings of the
codes $C_r \subset A^{n_r}$, that is, of viewing elements of each language as matrices
so that the concatenation operation of successive words can only happen for matrices
that has the same row lengths, while in the case of the spectrum of the maximal
abelian subalgebra one does not take the embedding into account so that all concatenations
of words in the languages defined by the codes $C_r$ are possible and one obtains a
larger set.

\medskip

{\bf 6.7. Quantum statistical mechanics above and below the asymptotic bound.}
We have seen in \S 4 how to associate a quantum statistical mechanical system
to an individual code. We also know from Theorem 2.10 that code points have
multiplicities: in particular, code points that lie below the asymptotic bound
have infinite multiplicity, while isolated codes, which lie above the asymptotic bound
have finite multiplicity. In terms of quantum statistical mechanical systems, it
is therefore more natural to fix a code point $(R,\delta)$ and construct an
algebra with time evolution $(TO_{(R,\delta)},\sigma)$ which does not depend
on choosing a code $C$ representing the code point, but allowing for all 
representatives simultaneously. This can be done in the same way we used 
in \S 6.4 for limit points. Namely, we let $TO_{(R,\delta)}$ be the Toeplitz algebra
with generators the elements in the union of all codes $C$ with parameters
$(R,\delta)$. This will be isomorphic to 
a finite rank Toeplitz algebra $TO_N$ for isolated codes and isomorphic to
the infinite Toeplitz algebra $TO_\infty$ in the case of code points that lie
below the asymptotic bound.  Similarly, we can consider the fractal set
given by the union of the $S_C$ for all the representative codes with fixed
$(R,\delta)$. In this case all these sets have the same Hausdorff dimension
equal to $R$, but in the case of isolated codes they are obtained as a finite
union and therefore they admit a uniform self-similar probability measure,
the $R$-dimensional Hausdorff measure, while in the case of the points
below the asymptotic bound one can construct non-uniform probability
measure using the same method we described in \S 6.2 for limit points.
We can use potentials as in (6.4) to construct such measures.
This in turn induces a time evolution on $TO_{(R,\delta)}$ of the form
$$ \sigma_t(T_a) = e^{it \lambda_a}\, T_a. $$
In this way, the properties of the quantum statistical mechanical system
associated to a code point $(R,\delta)$ reflect the difference between
point above or below the asymptotic bound.

\bigskip

\centerline{\bf 7. The asymptotic bound as a phase diagram.}

\medskip

The goal of this section is to extend the construction of quantum statistical mechanical systems
from the case of individual codes $C$ to families of codes in such a way as
to obtain a description of the asymptotic bound $R=\alpha_q(\delta)$ as a 
phase transition curve in a phase diagram.

\medskip

{\bf 7.1. Variable temperature KMS states.}
We begin by giving here a generalization of the usual notion of KMS states, which
we refer to as {\it variable temperature KMS states} and which will be useful in
our example. This is similar to the notion of ``local KMS states" considered, for
instance, in \cite{Acca} in the context of out of equilibrium thermodynamics, as well as 
in the context of information theory in [InKoO], though definition we give here is more
general. We formulate it first in the case of an arbitrary algebra of observables and we
then specialize it to the case of families of codes.
\medskip

{\bf 7.7.1. Definition.} {\it
Let $\Cal{B}$ be a unital $C^*$-algebra and let $\Cal{X}$ be a parameter space, assumed
to be a (compact Hausdorff) topological space, together with an assigned continuous 
function $\beta : \Cal{X} \to \R_+$. For $t\in C(\Cal{X},\R)$, let $\sigma_t \in \r{Aut}(\Cal{B})$ be
a family of automorphisms satisfying $\sigma_{t_1+t_2}=\sigma_{t_1} \circ \sigma_{t_2}$.
A KMS$_\beta$ state for $(\Cal{B},\sigma)$ is a continuous linear functional $\varphi:\Cal{B} \to \C$
with $\varphi(1)=1$ and $\varphi(a^* a)\geq 0$ for all $a\in \Cal{B}$, and such that, for all
$a,b \in \Cal{B}$ there exists a function $F_{a,b}(z)$, for $z: \Cal{X} \to \C$, with the property
that the function $F_{a,b}(z(\alpha))$ for any fixed $\alpha\in \Cal{X}$ and varying $z\in C(\Cal{X},\C)$
is a holomorphic function of the complex variable $z(\alpha)\in I_{\beta(\alpha)}$, where
$$ I_{\beta(\alpha)} = \{ z\in \C \,|\, 0 < \Re(z) < \beta(\alpha) \}, $$
and extends to a continuous function on the boundary of $I_{\beta(\alpha)}$ with
$$ F_{a,b}(t(\alpha)) = \varphi(a \sigma_{t(\alpha)}(b)), \ \ \ \text{ and } \ \ \
F_{a,b}(t(\alpha)+i\beta(\alpha)) =\varphi(\sigma_{t(\alpha)}(b) a), $$
where $t(\alpha)=z(\alpha)|_{\Re(z(\alpha))=0}$.}

\medskip
{\bf Example.} In the case where the parameter space is a finite set of points,
say $\Cal{X}=\{ 1, \ldots, N \}$ one finds that $\sigma_t$ is an action of $\R^N$ by
automorphisms and the variable temperature KMS condition gives a functional
such that $\varphi(a b)=\varphi(\sigma_{i\beta}(b)a)$, with $\beta=(\beta_1,\ldots,\beta_N)$.
The partition function, correspondingly, is a function $Z(\beta_1,\ldots,\beta_N)=\r{Tr}(e^{-\langle \beta, H\rangle})$, where $H=(H_k)$ implements the time evolution $\sigma_t$ in the sense that
$$ \pi(\sigma_t (a)) = e^{i\langle t, H \rangle} \pi(a) e^{-i \langle t, H \rangle}, $$
in a given Hilbert space representation $\pi$ of $\Cal{B}$.

\medskip
We are interested in the case where the algebra is itself a tensor product over
the parameter space, and the resulting $C^*$-dynamical system is also a
tensor product. Namely, we have $\Cal{B} = \otimes_{\alpha} \in \Cal{X} \Cal{B}_{\alpha}$
with $\sigma_t = \otimes_\alpha \sigma_{t(\alpha)}$ and a representation 
$\pi=\otimes_\alpha \pi_\alpha$ on a product $\Cal{H}=\otimes_\alpha \Cal{H}_\alpha$, 
with a Hamiltonian $H=\otimes_\alpha H_\alpha$ generating the time evolution, 
namely so that on $\Cal{H}_\alpha$ one has
$$ \pi_\alpha (\sigma_{t(\alpha)}(a_\alpha))=e^{i t(\alpha) H_\alpha} \pi_\alpha(a_\alpha) 
e^{-it(\alpha) H_\alpha}. $$
Then for a given $\beta:\Cal{X} \to \R_+$, a state $\varphi=\otimes_\alpha \varphi_\alpha$ is a 
KMS$_\beta$ state iff the $\varphi_\alpha$ are KMS$_{\beta(\alpha)}$ states for the
time evolution $\sigma_{t(\alpha)}$.  We assume here that $\Cal{X}$ is a discrete set and
that the $C^*$-algebras $\Cal{B}_\alpha$ are nuclear so that tensor products over finite
subsets of $\Cal{X}$ are unambiguously defined and the product over $\Cal{X}$ is obtained
as direct limit, as in Proposition 7 of [BoCo]. 

\medskip

{\bf 7.2. Phase transitions for families of codes.}
We consider approximations to the curve $R=\alpha_q(\delta)$ by 
families of $N$ points $(\delta_j,R_j)$ that are code points, that is,
for which there exists a code $C_j$ with $k_j/n_j = R_j$ and $d_j/n_j =\delta_j$.
To such a collection of points we associate a quantum statistical mechanical system
that is the tensor product of the systems associated to each code $C_j$, with
algebra of observables $\Cal{A} =\otimes_j TO_{C_j}$ and with the
dynamics given by $\sigma: \R^N \to \r{Aut}(\Cal{A})$, with $\sigma_t=\otimes_j \sigma_{t_j}$,
where $\sigma_{t_j}$ is the time evolution on $TO_{C_j}$ given by
$$ 
\sigma_{t_j}(S_a) = q^{it n_j} S_a. 
$$

\medskip
{\bf 7.2.1. Lemma.} {\it
Let $(\Cal{A},\sigma)$ be the product system described above, for a collection $C_j$
of codes, with $j=1,\ldots, N$. Then for any given $\beta=(\beta_1,\ldots,\beta_j)$ 
there is a unique KMS$_\beta$ state on $(\Cal{A},\sigma)$, which is given by the product
$\varphi_\beta=\otimes_j \varphi_{\beta_j}$ of the unique KMS$_{\beta_j}$ states
on the algebras $TO_{C_J}$. For $\beta$ in the region $\beta_j > R_j$, the KMS
state is of type I$_\infty$. The partition function is the product of the partition functions 
of the individual systems.}

\smallskip

{\bf Proof.} 
The product state $\varphi_\beta=\otimes_j \varphi_{\beta_j}$ is a KMS$_\beta$ state
for $(\Cal{A},\sigma)$ with $\beta=(\beta_1,\ldots,\beta_j)$. The uniqueness for the tensor
product state follows from an argument similar to the one used in Proposition 8 of
[BoCo], adapted to our more general notion of KMS state. It suffices in fact to observe
that, if $\varphi$ is a KMS$_\beta$ state with $\beta=(\beta_1,\ldots,\beta_j)$ on the
product $\Cal{A}=\otimes_j TO_{C_j}$, then for fixed $a_1,\ldots, a_{j-1}, a_{j+1},\ldots, a_N$, 
the functional 
$$ \varphi_{a_1\otimes \cdots \otimes a_{j-1}\otimes a_{j+1} 
\otimes \cdots \otimes a_N}(a_j)=
\varphi(a_1\otimes \cdots \otimes a_N) $$
is a KMS$_{\beta_j}$ state on $TO_{C_j}$, by the same argument used in the ordinary case.

The Hamiltonian $H_j$ generating the time evolution $\sigma_{t_j}$ on the algebra
$TO_{C_j}$ has eigenvalues $m n_j \log q$, with integers $m\geq 0$, with multiplicities
$q^{m k_j}$, and partition function
$$ Z(\beta_j)=\r{Tr}(e^{-\beta_j H_j}) = \sum_m q^{(R_j-\beta_j)n_j m} =
(1-q^{(R_j-\beta_j)n_j})^{-1}. $$
The partition function for the product system is then
$$ Z(\beta_1,\ldots,\beta_N)= \r{Tr}(e^{-\sum_j \beta_j H_j}) 
= \sum_{m=(m_1,\ldots, m_N)}  q^{\sum_j (R_j-\beta_j)n_j m_j} $$ $$ =
\prod_j (\sum_{m_j} q^{(R_j-\beta_j)n_j m_j}) = \prod_j (1-q^{(R_j-\beta_j)n_j})^{-1} =
\prod_j Z(\beta_j). $$
It converges in the domain of $\R^N$ determined by the conditions $\beta_j > R_j$.

\smallskip

This finishes the proof.

\medskip

To further refine the picture described above, we consider quantum statistical
mechanical systems associated to families of codes approximating a limit
point in the closure of the code domain.

\smallskip
As before, let $\Cal{C} =\{ C_r \}$ be a family of codes with $k_r/n_r \nearrow R$ and
$d_r/n_r \nearrow \delta$. We consider again the union $\cup_r C_r$
and the corresponding Toeplitz algebra $TO_{\cup_r C_r}$. 
On the fractal $S_R =\cup_r S_{C_r}$ 
of Hausdorff dimension $\dim_H (S_R)=R$, consider a potential 
$W_\beta(x) = e^{-\beta \lambda_{x_1}}$, such that, when $\beta=R$ it satsifies the
Keane condition
$$ \sum_{a\in \cup_r C_r} e^{- R \lambda_a} =1. $$
We consider then the time evolution on $TO_{\cup_r C_r}$ given by
$$ \sigma^W_t(T_a) = e^{it \lambda_a} T_a. $$
In the representation of $TO_{\cup_r C_r}$ on its Fock space, this time evolution is generated
by a Hamiltonian 
$$ H \epsilon_w = (\lambda_{w_1}+\cdots+ \lambda_{w_m}) \epsilon_w, $$
for $w=w_1\cdots w_m$ with $w_i \in \cup_r C_r$. 
This has partition function
$$ Z_{\Cal{C}}(\beta) = \sum_m \sum_{w\in \Cal{W}_{\cup_r C_r,m}} e^{-\beta (\lambda_{w_1}+\dots+ \lambda_{w_m}) } =\sum_m \left(\sum_{a\in \cup_r C_r} e^{-\beta \lambda_a}\right)^m. $$
If we introduce the notation
$$ \Lambda(\beta):= \sum_{a\in \cup_r C_r}  e^{-\beta \lambda_a} , $$
we have $\Lambda(R)=1$ and, for $\beta > R$, $\Lambda(\beta)<1$, while for
$\beta< R$ one has $\Lambda(\beta) >1$, which becomes possibly divergent after some
critical value $\beta_0 < R$.  Thus, the partition function for the system
$(TO_{\cup_r C_r},\sigma^{W})$ is
$$ 
Z_{\Cal{C}}(\beta) = \sum_m \Lambda(\beta)^m = (1-\Lambda(\beta))^{-1}, 
$$
convergent for $\beta > R$, with a phase transition at $\beta =R$. 
The same argument of Proposition 4.7.3 can be extended to this
case to show the existence at all $\beta >0$ of a unique KMS$_\beta$ state,
which is of type I$_\infty$ below the critical temperature and is given by a 
residue at the critical temperature.

\medskip

One can then consider approximations of the curve $R=\alpha_q(\delta)$ by points
$(R_{j,N},\delta_{j,N})$ in $U_q$, for $j=1,\ldots,N$.
To each of these points one associates a
quantum statistical mechanical system constructed as above using a family 
$\Cal{C}_{j,N}=\{ C_{r_{j,N}} \}$ of codes approximating the limit point 
$(\delta_{j,N}, R_{j,N})$ with the time evolution $\sigma^{W_{j,N}}$ described above
on the algebra $TO_{\Cal{C}_{j,N}}$. By taking the product of these systems
one can form a system with variable temperature KMS states with phase
transition at $\beta_{j,N}= R_{j,N}\leq \alpha_q(\delta_{j,N})$. This can be extended
to the case of a countable dense set of points below the curve $R=\alpha_q(\delta)$
and the corresponding countable tensor product system. 

\medskip

It would be interesting to extend this type of tensor product construction for
families of algebras associated to codes to a version that corresponds to a
``system with interaction" more like the Bost--Connes algebra.

\bigskip

\bigskip

\centerline{\bf References}

\medskip

[Ac] L.~Accardi, K.~Imafuku, {\it Dynamical detailed balance and local KMS 
condition for non-equilibrium states}, International Journal of Modern Physics B, Vol.18
(2004), no. 4--5, 345--467.

\smallskip

[BoCo] J.B.~Bost, A.~Connes, {\it Hecke algebras, Type III factors and phase
transitions with spontaneous symmetry breaking in Number Theory}, Selecta Math. (New Ser.)
Vol. 1 (1995), no. 3, 411--457.

\smallskip

[Co1] A.~Connes. {\it A survey of foliations and operator algebras}. Proc. Sympos. Pure Math.
Vol. 38, Part I (1982), 85--115.

\smallskip

[Co2] A.~Connes. {\it Une classiÞcation des facteurs de type III}. Ann. Sci. 
\'Ecole Norm. Sup. (4) 6 (1973),  133--252. 

\smallskip

[CoCoMar] A.~Connes, C.~Consani, M.~Marcolli, {\it Noncommutative geometry and motives: the thermodynamics of endomotives}, Advances in Mathematics, Vol. 214 (2007), no. 2, 761--831.

\smallskip

[Cu1] J.~Cuntz, {\it Simple $C^*$--algebras generated by isometries}. 
Commun. Math. Phys. 57 (1977), 173--185. 

\smallskip

[Cu2] J.~Cuntz. {\it K--theory for certain $C^*$--algebras}. Ann. Math. 113
(1981), 181--197.

\smallskip

[DutJor] D.E.~Dutkay, P.E.T.~Jorgensen. {\it Iterated function systems, 
Ruelle operators, and invariant projective measures}. Math. Comp. 75 (2006), 1931--1970.

\smallskip

[Eilen] S.~Eilenberg, {\it Automata, languages, and machines}, Vol.~A, Academic Press,
1974. 

\smallskip

[Ex] R.~Exel, {\it A new look at the crossed-product of a $C^*$-algebra by an 
endomorphism},  Ergodic Theory and Dynamical Systems, Vol. 23 (2003), 1733--1750.

\smallskip

[Fal] K.~Falconer. {\it Fractal geometry}. Wiley, 1990.

\smallskip

[Fow] N.~J.~Fowler. {\it States of Toeplitz--Cuntz algebras.} J.~Operator Theory, 42 (1999), no. 1,
121--144. arxiv:funct-an/9702012

\medskip

[InKoO] R.S.~Ingarden, A.~Kossakowski, and M.~Ohya, {\it Information Dynamics and Open systems}, Kluwer Accademic Publishers, 1997.

\smallskip

[KiKu] A.~Kishimoto, A.Kumjian. {\it Simple stably projectionless 
$C^*$--algebras arising as crossed products}. Can. J. Math., Vol. 48 (1996), 980--996.

\smallskip

[KuRe] A.~Kumjian, J.~Renault. {\it KMS states on $C^*$--algebras 
associated to expansive maps}. Proc. AMS, Vol. 134 (2006), 2067--2078. 

\smallskip

[LiVi] M.~Li, P.M.B.~Vit\'anyi, {\it An introduction to Kolmogorov
complexity and its applications}, 2nd edition, Springer, 1997.

\smallskip

[LinPh] H.~Lin, N.C.~Phillips. {\it Approximate unitary equivalence of 
homomorphisms from $O_\infty$}. J. Reine Angew. Math., 464 (1995), 173--186.

\smallskip

[Man] Yu.~I.~Manin, {\it What is the maximum number of points
on a curve over $\bold{F}_2$?} J. Fac. Sci. Tokyo, IA, Vol. 28 (1981),
715--720. 

\smallskip

[ManVla] Yu.~I.~Manin. S.G.~Vladut, {\it Linear codes and
modular curves}. J. Soviet Math., Vol. 30 (1985), 2611--2643.
\smallskip

[Mar] M.~Marcolli, {\it Cyclotomy and endomotives}, 
p-Adic Numbers, Ultrametric Analysis and Applications, Vol.1 (2009), no. 3, 217--263.

\smallskip

[MarPa] M.~Marcolli, A.~M.~Paolucci, {\it Cuntz--Krieger algebras
and wavelets on fractals}. arXiv:0908.0596.
\smallskip

[Rya1] B.Ya.~Ryabko, {\it Noiseless coding of combinatorial sources}, Problems in Information
Transmission, 22 (1986), 170--179. 

\smallskip

[Rya2] B.Ya.~Ryabko, {\it Coding of combinatorial sources and Hausdorff dimension},
Soviet Math. Doklady, 30 (1984), no. 1, 219--222.

\smallskip

[Sta1] L.~Staiger, {\it The Kolmogorov complexity of infinite words}. 
Theoret. Comput. Sci. 383 (2007), no. 2-3, 187--199.

\smallskip

[Sta2] L.~Staiger, {\it Constructive dimension equals Kolmogorov complexity}. 
Inform. Process. Lett. 93 (2005), no. 3, 149--153.

\smallskip

[Sta3]  L.~Staiger, {\it Kolmogorov complexity and Hausdorff dimension}, Inform. and Comput. 103 (1993), 159--194.

\smallskip

[TsfaVla] M.~A.~Tsfasman, S.~G.~Vladut, {\it Algebraic--geometric
codes}, Kluwer, 1991.

\smallskip

[UShe] V.A.~Uspensky, A.Shen, {\it Relations between varieties of Kolmogorov complexity},
Math. Systems Theory 29 (1996), 271--292.

\smallskip

[ZvoLe] A.K.~Zvonkin, L.A.~Levin, 
{\it The complexity of finite objects and the basing of the concepts of information and randomness on the theory of algorithms}, Russian Math. Surveys 25 (1970), no. 6, 83--124.

\bigskip

\bigskip

YURI I.~MANIN, 

{\it Max Planck Institute for Mathematics, Bonn and

Mathematics Department, Northwestern University }

manin\@mpim-bonn.mpg.de

\bigskip

MATILDE MARCOLLI, 

{\it Department of Mathematics, California Institute of Technology and

Max Planck Institute for Mathematics, Bonn}

matilde\@caltech.edu

\enddocument